\documentclass[a4paper, 11pt, leqno]{article}

\usepackage[utf8]{inputenc}		% for use of umlauts in the BibTeX-file
\usepackage[english]{babel}
\addto{\captionsenglish}{}
\usepackage{amsmath}
\usepackage{amsfonts}
\usepackage{amsthm}
\usepackage{amssymb}
\usepackage{mathrsfs}			% for script math
\usepackage{mathtools}			% for matrices with aligned columns, and correct vertical spacing of adjacent inf and sup.
\usepackage{array}				% for displaystyle in arrays
\usepackage{float}
\usepackage[shortlabels]{enumitem}
\usepackage{tikz}						% for commutative diagrams
\usepackage{pict2e}				% for arbitrary slopes in the \line command
\usepackage{bbm}				%for boldface 1
\usepackage[a4paper]{geometry}	% for better margins
\usepackage{hyperref,bookmark}
\usepackage{caption}

\pdfoutput=1

\providecommand{\noopsort}[1]{} %for tussenvoegsel-sorting in bibtex: \noopsort

\allowdisplaybreaks		% to allow aligned equations to spread across multiple pages
\newcommand{\IN}{\mathbb{N}}
	\newcommand{\N}{\mathbb N}
\newcommand{\IZ}{\mathbb{Z}}
	\newcommand{\Z}{\mathbb Z}

\newcommand{\IR}{\mathbb{R}}
	\newcommand{\R}{\mathbb R}
\newcommand{\IC}{\mathbb{C}}
	\newcommand{\C}{\mathbb C}
\newcommand{\IT}{\mathbb{T}}
	\newcommand{\T}{\mathbb T}
\renewcommand{\S}{\mathcal{S}}
\renewcommand{\O}{\mathcal{O}}
\newcommand{\mR}{\mathcal{R}}
\newcommand{\mB}{\mathcal{B}}
\newcommand{\Cr}{C_\mR}
\newcommand{\Sr}{\S_\mR}
\newcommand{\Wr}{\mathcal{W}^0_\mR}
\newcommand{\mN}{\mathcal{N}}

\newcommand{\cH}{\mathcal{H}}
\newcommand{\unit}{\mathbf{1}}
\newcommand{\set}[2]{\{#1 \colon #2 \}}
\newcommand{\inp}[2]{\langle #1, #2 \rangle}
\newcommand{\hatotimes}{\mathbin{\hat{\otimes}}}
\newcommand{\Weylalg}[1]{\mathcal{W}^0(#1)}
\newcommand{\Teunalg}[1]{\mathcal{W}^0_{\mathcal{R}}(#1)}
\newcommand{\Id}{\text{\normalfont Id}}

\newcommand{\supp}{\text{\normalfont supp}}

\renewcommand{\mid}{~\middle|~}
\newcommand{\QW}{\mathcal{Q}_\hbar^W}

\newcommand{\spn}{\textnormal{span}}
\newcommand{\dom}{\operatorname{dom}}
\newcommand{\vect}[2]{\begin{pmatrix}#1\\#2\end{pmatrix}}
\newcommand{\norm}[1]{\left\| #1 \right\|}

\newcommand{\supnorm}[1]{\norm{ #1 }_\infty}
\newcommand{\dt}{\Delta t}

\theoremstyle{plain}
\newtheorem{thrm}{Theorem}
\newtheorem{lem}[thrm]{Lemma}
\newtheorem{prop}[thrm]{Proposition}

\theoremstyle{definition}
\newtheorem{defi}[thrm]{Definition}

\newtheorem{rem}[thrm]{Remark}

\newenvironment{myenum}
{\begin{enumerate}[{(}1{)}]}
{\end{enumerate}}

\newenvironment{proofenum}
{
\begin{proof}

\noindent
\begin{enumerate}[label={\bf (\arabic*) \:}, wide, labelwidth=!, labelindent=0pt, parsep=0pt]
}
{
\qedhere
\end{enumerate}
\end{proof}
}

\newenvironment{claimenum}[1]
{
\vspace{-8pt}
\begin{enumerate}[label={\it (\alph*) \:}, topsep=12pt, partopsep=0pt, leftmargin=\parindent, itemindent=\parindent, labelwidth=\itemindent, labelsep=0em, itemsep=0em, parsep=0pt, align=left, start=#1]
\em 
}
{
\end{enumerate}
\vspace{-8pt}
}

\begin{document}

%{\small 
%\tableofcontents
%\thispagestyle{plain}
%}

\newpage

% begin main part
\def\mycmd{1}
\ifx\mycmd\undefined
	\documentclass[a4paper, 11pt, leqno]{article}
	
	\begin{document}
\fi

\noindent

\title{Classical and quantised resolvent algebras for the cylinder}
\author{{T.D.H. van Nuland\thanks{t.vannuland@math.ru.nl},\qquad R. Stienstra\thanks{r.stienstra@math.ru.nl}}\\
Radboud University Nijmegen, Heyendaalsweg 135,\\ 6525 AJ Nijmegen, The Netherlands}

\maketitle

\begin{abstract}

\noindent
Buchholz and Grundling ({\em Comm.
Math. Phys.}, 272, 699--750, 2007) introduced a C$^\ast$-algebra called the resolvent algebra as a canonical quantisation of a symplectic vector space, and demonstrated that this algebra has several desirable features.
We define an analogue of their resolvent algebra on the cotangent bundle $T^*\T^n$ of an $n$-torus by first generalizing the classical analogue of the resolvent algebra defined by the first author of this paper in earlier work ({\em J. Funct. Anal.},
277, 2815--2838, 2019), and subsequently applying Weyl quantisation.
We prove that this quantisation is almost strict in the sense of Rieffel and show that our resolvent algebra shares many features with the original resolvent algebra.
We demonstrate that both our classical and quantised algebras are closed under the time evolutions corresponding to large classes of potentials.
Finally, we discuss their relevance to lattice gauge theory.
\end{abstract}

\section{Introduction}
\label{subsec:classical_resolvent_algebra_introduction}

\noindent
	Much of modern physics concerns the search for and examination of quantum versions of known classical theories. Examples include quantum statistical mechanics, quantum field theory, and quantum gravity.
Showing that a classical theory is indeed the limit of the quantum theory at hand can be done at various levels of rigour.
The most precise way to establish this limit is by strict deformation quantisation, where one `quantises' a classical (commutative) Poisson algebra into a quantum (noncommutative) C*-algebra \cite{rieffel89, landsman98} (cf. \cite[p. 5]{hawkins08} for an overview of the various definitions in the literature).

	Only few pairs of a classical and a quantum C*-algebra are known to connect in this rigorous fashion \cite{ bieliavsky15, landsman98, rieffel90, rieffel93}, and each has its merits and drawbacks.
In particular, when taking the torus as a configuration space, we found the known examples too limited in certain respects. Hence, in this paper, we define a quantum observable algebra on the torus, i.e. a C*-algebra $A_\hbar\subseteq B(L^2(\T^n))$, which satisfies the following properties:
\begin{enumerate}[label=P\arabic{*}:, ref=P\arabic{*}]
	\item The algebra $A_\hbar$ has a classical counterpart $A_0$ and can be obtained from this commutative algebra through (strict) quantisation.\label{property:quantisation}
	\item The algebra $A_\hbar$ is closed under the time evolution associated to the potential $V$ for each $V \in C(\IT^n)$.
The classical analogue $A_0$ satisfies a similar condition.\label{property:time}
	\item The classical and quantum algebras associated to a given system are both sufficiently large to accommodate natural embeddings of the respective algebras corresponding to their subsystems.\label{property:direct}
	\item The algebras $A_0$ and $A_\hbar$ are the smallest C$^\ast$-algebras satisfying the previous conditions whilst containing the algebra $C_0(T^\ast \mathbb{T}^n)$ and its quantisation $\mathcal K(L^2(\IT^n))$, respectively.
	\label{property:small}
\end{enumerate}

\noindent
An observable algebra satisfying only \ref{property:quantisation}, \ref{property:time} and \ref{property:small} has long been known, namely $\mathcal K(L^2(\IT^n))$, the compact operators on $L^2(\IT^n)$, with $C_0(T^*\IT^n)$ as its classical limit (cf. \cite{landsman98}, in particular sections II.3.4, III.3.6 and III.3.11).
We now sketch how the need for \ref{property:direct} arises in quantum lattice gauge theory.
Although a significant portion of this introduction is dedicated to this argument, it is presented here in condensed form; more details can be found in \cite[section 5.1]{stienstra19}.

\paragraph{Lattice gauge theory} In the Hamiltonian lattice gauge theory by Kogut and Susskind \cite{kogut75}, one approximates a time-slice of spacetime by a finite `lattice', or more accurately, an oriented graph $\Lambda$.
The elements of the set of vertices $\Lambda^0$ are points in the time slice, while the set of oriented edges $\Lambda^1$ are paths between these points.
A gauge field corresponding to some connection on a principal fibre bundle over spacetime with structure group some Lie group $G$ is approximated by the parallel transport maps along the edges of $\Lambda$.
After choosing a trivialisation of the restriction of the principal fibre bundle to $\Lambda^0$, the set of all possible parallel transporters can be identified with $G^{\Lambda^1}$; this is the configuration space of the Hamiltonian lattice gauge theory, and it carries a natural action of $G^{\Lambda^0}$ (endowed with the obvious group structure).
This group is the analogue of the set of gauge transformations.

Let us take a brief moment to comment on the choice of the structure group $G$ of the gauge theory.
Lattice gauge theory was originally introduced by Wilson \cite{wilson74} to explain the phenomenon of quark confinement in the context of the gauge theory known as {\em quantum chromodynamics (QCD)}.
The underlying structure group of QCD is $\text{\normalfont SU}(3)$, hence the corresponding configuration space is evidently not a torus and therefore lattice QCD is outside of the scope of this article.
However, it is worth noting that the structure group of electromagnetism is $\IT$, so the results in this paper may be applied to corresponding lattice gauge theories, or perhaps serve as a stepping stone towards an analogous construction that can be applied to lattice QCD.

We now return to the argument.
The Hilbert space of the corresponding quantum lattice gauge theory is $\cH = L^2(G^{\Lambda^1})$, where $G^{\Lambda^1}$ is endowed with the normalised Haar measure.
The field algebra of the system is some C$^\ast$-algebra $A_\Lambda$ that is represented on $\cH$, from which the observable algebra can be obtained by applying a reduction procedure with respect to the gauge group (cf. \cite{kijowski04, stienstra18}).
The observable algebra is accordingly represented on the set of elements of $\cH$ that are invariant under gauge transformations.
Since the distinction between field and observable algebras is irrelevant with regard to the issue that motivates the present investigation -- the embedding maps take the same form in both cases -- we will continue to refer to $A_\Lambda$ as the observable algebra in what follows.

In the context of lattice gauge theory, one is interested in constructing an algebra of the continuum system from the above algebras $A_\Lambda$.
This is done by considering direct systems of lattices, and we are naturally led to consider the following situation.
Suppose that $\Lambda_1$ and $\Lambda_2$ are both lattices approximating a time slice, and that $\Lambda_2$ is a better approximation than $\Lambda_1$, i.e., $\Lambda_1^0 \subseteq \Lambda_2^0$, the graph $\Lambda_2$ contains more paths than $\Lambda_1$, and each edge in $\Lambda^1$ can be written as a concatenation of paths in $\Lambda^2$; for a precise definition, we refer to \cite{arici18}.
We should then be able to find a corresponding embedding map $A_{\Lambda_1} \hookrightarrow A_{\Lambda_2}$.
The embedding map takes a simple form if $\Lambda_2$ is obtained from $A_{\Lambda_1}$ by only adding edges: in that case, we have $\cH_2 = \cH_1 \hatotimes \cH_1^c$, where $\cH_1^c = L^2(G^{\Lambda_2^1 \backslash \Lambda_1^1})$, and the embedding is given by the restriction of the map
\begin{equation*}
B(\cH_1)\rightarrow B(\cH_2) \cong B(\cH_1) \hatotimes B(\cH_1^c)\,,\qquad a \mapsto a \otimes \mathbf{1}\,,
\end{equation*}
to $A_{\Lambda_1}$, where $\mathbf{1}$ denotes the identity on $\cH_1^c$, and $\hatotimes$ denotes the von Neumann algebraic tensor product.

A first guess for the observable algebras of the two quantum systems could be $\mathcal K(\cH_1)$ and $\mathcal K(\cH_2)$, the algebras of compacts.
However, except in trivial cases, the Hilbert space $\cH_1^c$ will be infinite-dimensional, which means that $a\otimes \mathbf{1}$ will not be a compact operator.
Thus the algebra $\mathcal K(\cH_2)$ is too small to accommodate these embeddings.
This problem was already noticed by Stottmeister and Thiemann in  \cite{stottmeister16}.
In an earlier paper \cite{arici18} on Hamiltonian lattice gauge theory coauthored by one of the authors of the present article, the above problem was not encountered since different embedding maps were used.
There are nevertheless good reasons to believe that the embedding maps used in \cite{stottmeister16} are the correct ones, though we will not elaborate on them here, and refer to \cite[Chapter 8]{stienstra19} instead.
The argument presented there is not specific to lattice gauge theory, but can be made for any physical system that is comprised of smaller subsystems.

Another guess for the observable algebra of the composite system could be the one generated by the embedded algebras of all subgraphs, as is done in \cite{grundling17}. However, this raises questions about regulator independence of this procedure in situations where one takes limits corresponding to an infinite volume or continuum limit of a collection of systems parametrised by a cutoff.
As this problem is beyond the scope of the present article, we will refer the reader to the discussion in \cite[section 5.1]{stienstra19}.
The main point is that there is ample reason to try to solve the problem through an appropriate choice of algebras, i.e., algebras that satisfy \ref{property:direct}.

\noindent 
\paragraph{The resolvent algebra on $\R^n$} In the case where the configuration space is $\R^n$, there already exists an algebra satisfying \ref{property:quantisation}, \ref{property:time}, \ref{property:direct} and \ref{property:small}, namely the resolvent algebra $\mR(\R^{2n},\sigma_n)$.
The resolvent algebra $\mathcal{R}(X,\sigma)$ on a symplectic vector space $(X,\sigma)$ is a C$^\ast$-algebra that was originally introduced by Buchholz and Grundling in \cite{buchholz07}, and subsequently studied in greater detail in \cite{buchholz08} and \cite{buchholz14} by the same authors.
Before we adapt this algebra to the case of $T^*\T^{n}$ instead of $\R^{2n}$ as its underlying phase space, let us recall the main idea behind the construction of the resolvent algebra.

The resolvent algebra is constructed as the completion of a $^\ast$-algebra with respect to a certain C$^\ast$-seminorm \cite[Defintion 3.4]{buchholz08}; the $^\ast$-algebra is defined in terms of generators and relations.
To each pair $(\lambda, f) \in (\IR \backslash \{0\}) \times X$, a generator $\mathcal{R}(\lambda, f)$ is associated.
Such a generator is thought of as the resolvent (depending on $\lambda$) corresponding to some unbounded operator $\phi(f)$ associated to the vector $f$, where $\phi$ denotes a linear map from $X$ to a space of operators on a dense subspace of a Hilbert space on which $\mathcal{R}(X, \sigma)$ can be represented faithfully.

For example, suppose that $(X, \sigma)$ is $\IR^2$ endowed with the standard symplectic form.
Then $\mathcal{R}(X, \sigma)$ admits a faithful representation on $L^2(\IR)$ such that the unbounded operators corresponding to the vectors $(1,0)$ and $(0,1)$ are the standard position and momentum operators respectively (up to a factor of $\hbar$ in the latter case), see \cite[Corollary 4.4 and Theorem 4.10]{buchholz08}.
Both of these unbounded operators can be defined on the (invariant) dense subspace $C^\infty_c(\IR)$, on which they are essentially self-adjoint.

For each $f \in X$, the generator $\mathcal{R}(\lambda, f)$ is mapped to the bounded operator $(i\lambda \unit - \phi(f))^{-1}$; in particular, taking $f = 0$, we see that $\mathcal{R}(X, \sigma)$ is unital.
The relations defining the $^\ast$-algebra from which the resolvent algebra is constructed serve to encode the fact that $\mathcal{R}(\lambda, f)$ behaves like the resolvent of the unbounded operator $\phi(f)$, as well as the linearity of $\phi$.
Last but not least, the canonical commutation relations (CCR) are introduced by the defining relations of $\mathcal{R}(X, \sigma)$ in which the symplectic form appears, thereby justifying the term ``canonical quantum systems'' in the title of \cite{buchholz08}.

The resolvent algebra is not the only approach to the reformulation of the CCR in a framework based on bounded operators; another is obtained through exponentiation of the unbounded operators of interest, leading to the Weyl form of the CCR and the Weyl algebra.
There is a bijection between certain classes of representations of these two algebras \cite[Corollary 4.4]{buchholz08}.
In particular, generators of the resolvent algebras can be expressed in terms of generators of the Weyl algebra by means of the Laplace transform, as is done in \cite{buchholz07}.
By changing the representation in that definition to the usual representation on $L^2(\IR)$ of the Weyl algebra on $\IR^2$, one obtains the representation mentioned earlier.

Buchholz and Grundling note that their resolvent algebra has some desirable qualities not shared by the Weyl algebra, such as the presence of observables corresponding to bounded functions in regular representations.
Furthermore - and this is particularly relevant for this paper - the resolvent algebra associated to $\IR^2$ endowed with the standard symplectic form is closed under (quantum) time evolution for a much larger class of Hamiltonians than the Weyl algebra (cf. \cite[Proposition 6.1]{buchholz08}).
The authors explain this as a consequence of the fact that their resolvent algebra contains resolvents of many Hamiltonians.
Moreover, Buchholz has shown that the resolvent algebra is stable under dynamics in the context of oscillating lattice systems \cite{buchholz17} and nonrelativistic Bose fields \cite{buchholz18}. 

According to \cite[Theorem 5.1]{buchholz08}, for any symplectic vector space $(X,\sigma)$ and any decomposition $S \oplus S^\perp$ of $X$ into subspaces that are nondegenerate with respect to $\sigma$ (and were $S^\perp$ denotes the complement of $S$ with respect to $\sigma$), the resolvent algebra $\mathcal{R}(X, \sigma)$ naturally contains a copy of $\mathcal{R}(S, \sigma) \hatotimes \mathcal{R}(S^\perp, \sigma)$; with respect to corresponding faithful representations of these three resolvent algebras, the embeddings of $\mathcal{R}(S, \sigma)$ and $\mathcal{R}(S^\perp, \sigma)$ are given by the analogues of the aforementioned embedding map for lattice gauge theory.
Here, $\hatotimes$ denotes any C*-norm (nuclearity of the resolvent algebra is shown in \cite{buchholz14}), and $\sigma$ by abuse of notation denotes the symplectic form on $X$, as well as its restrictions to $S$ and $S^\perp$.\\

\noindent
We have seen how properties \ref{property:direct} and \ref{property:time} where shown by Buchholz and Grundling to hold for the resolvent algebra. Now for \ref{property:quantisation} the question is whether it arises as the strict quantisation of an algebra that can be considered the observable algebra of a classical system in the sense of Landsman, i.e., the C$^\ast$-algebra generated by the image of a dense Poisson subalgebra of the classical algebra under a quantisation map \cite{landsman98}.
This question was answered affirmatively by one of the authors of this paper in \cite{vnuland19}, where it is shown that in the case where $(X, \sigma)$ is $\IR^{2n}$ endowed with the standard symplectic form, there is a corresponding commutative C$^\ast$-algebra $C_{\mathcal{R}}(\IR^{2n})$, which is the C$^\ast$-subalgebra of $C_b(\IR^{2n})$ generated by functions of the form
\begin{equation*}
x \mapsto (i\lambda - x \cdot v)^{-1}, \quad
\lambda \in \IR \backslash \{0\}, \:
v \in \IR^{2n},
\end{equation*}
where $\cdot$ denotes the standard inner product.
Similar to the way in which the algebra $C_0(\IR^{2n})$ may be quantised into the compact operators on $L^2(\IR^n)$ by considering the dense Poisson subalgebra $\mathcal{S}(\IR^{2n})$ of Schwartz functions and defining Weyl or Berezin quantisation on them, an analogue of the space of Schwartz functions for $C_{\mathcal{R}}(\IR^n)$ is identified as follows.
First, for every linear subspace $V \subseteq \IR^n$, let $P_V$ denote the orthogonal projection onto $V$.
Then the space
\begin{equation*}
\mathcal{S}_{\mathcal{R}}(\IR^n)
:= \text{\normalfont span}_{\mathbb{\IC}} \set{g \circ P_V}{V \subseteq \IR^n \text{ is a subspace, } g \in \mathcal{S}(V)} \, ,
\end{equation*}
is defined.
It is readily seen that this is a dense Poisson subalgebra of $C_{\mathcal{R}}(\IR^n)$ that is closed under the $^\ast$-operation of complex conjugation.
The Weyl quantisation of $g \circ P_V$ is defined using the Fourier transform of $g$ as a function on $V$ \cite[section 3.2]{vnuland19}, but is otherwise equal to the definition of the Weyl quantisation of ordinary Schwartz functions on $\IR^n$.
It is then argued that the Weyl quantisation map admits a (unique) linear extension to $\mathcal{S}_{\mathcal{R}}(\IR^n)$.
Furthermore, it is shown that the images of $\mathcal{S}_{\mathcal{R}}(\IR^n)$ under Weyl and Berezin quantisation are both dense subspaces of $\mathcal{R}(\IR^{2n}, \sigma)$.
The resulting algebra $C_{\mathcal{R}}(\IR^{2n})$ is accordingly referred to as the {\em classical resolvent algebra on $\IR^{2n}$}.
As is shown in \cite{vnuland19}, these definitions are easily extended to spaces of functions whose domain is an inner product space of arbitrary dimension.

In addition to being the classical counterpart of the resolvent algebra as defined by Buchholz and Grundling, the classical resolvent algebra offers an interesting perspective on our earlier discussion on embeddings of observable algebras.
In some sense, $C_{\mathcal{R}}(\IR^n)$ is the smallest C$^\ast$-subalgebra of $C_b(\IR^n)$ that contains $C_0(\IR^n)$, whilst also containing its analogues associated to linear subspaces of $\IR^n$.
This may be formalised as follows.
Consider the category whose objects are finite-dimensional real vector spaces, and whose morphisms consist of projections of a vector space onto one of its subspaces.
Then there is a contravariant functor $C_b$ from this category to the category of C$^\ast$-algebras that maps an object $V$ to the space $C_b(V)$, and that maps morphisms to their pullbacks between these spaces.
It is now consistent with the definition of the classical resolvent algebra to define $C_{\mathcal{R}}$ as the smallest subfunctor of $C_b$ with the property that the image of every object $V$ contains $C_0(V)$.
Note that this implies that $C_\mathcal{R}(\IR^n)$ is unital, as it contains the embedding of $C_0(\{0\})$.
This makes precise in which sense \ref{property:small} holds for the resolvent algebras on $\IR^{2n}$.

\noindent
\paragraph{Resolvent algebras on the cylinder}
In this paper, we define an analogue of the resolvent algebra for the cotangent bundle $T^\ast \IT^n \cong \IT^n \times \IR^n$ of the $n$-torus $\IT^n$.
Our approach differs significantly from that of Buchholz and Grundling, in that we do not define it in terms of generators and relations.
Rather, we first identify a classical resolvent algebra $C_{\mathcal{R}}(T^\ast \IT^n)$ using ideas from \cite{vnuland19}, and indicate how this definition may be generalised.
We then give a concrete characterisation of $\Cr(T^*\T^n)$. Namely, identifying $T^*\T^n$ with $\T^n\times\R^n$, we prove that $\Cr(T^*\T^n)$ equals $C(\T^n)\!\hatotimes\! \Wr(\R^n)$, where $\Wr(\R^n)$ is the C*-algebra generated by the functions 
	\begin{align}\label{generators}x\mapsto 1/(i+x\cdot v)\quad\text{ and }\quad x\mapsto e^{ix\cdot v}\,,\quad\text{ for all }\quad v\in\R^n\,.
	\end{align}
In addition, we identify a dense $^\ast$-subalgebra $\mathcal{S}_{\mathcal{R}}(T^\ast \IT^n)\subseteq\Cr(T^*\T^n)$ carrying a natural Poisson structure.
The algebra is spanned by functions of the form $e_k\otimes h$, where $e_k[x]:=e^{2\pi ik\cdot x}$, and $h$ is a smooth function that is a product of an element of $\mathcal{S}_{\mathcal{R}}(\R^n)$ and a function of the form $x \mapsto e^{i \xi \cdot x}$ for some $\xi \in \IR^n$.

To define a quantum counterpart, we apply Weyl quantisation, making \ref{property:quantisation} integral to the definition of the (quantum) resolvent algebra on $T^*\T^n$.
Our Weyl quantisation map $\QW:\Sr(T^*\T^n)\rightarrow B(L^2(\T^n))$ is defined with an integral formula inspired by \cite{rieffel93}. When writing $\Cr(T^*\T^n)$ as a tensor product as above, $\QW$ can be characterised by the formula
\begin{align}\label{eq:formula QW intro}
	\QW(e_k\otimes h)\psi_l=h(\pi\hbar(k+2l))\psi_{k+l}\,,
\end{align}
where $e_k\otimes h\in\Sr(T^*\T^n)$, and $\psi_k$ is $e_k$ viewed as an element of $L^2(\mathbb{R}^n)$ for each $k \in \mathbb{Z}^n$.
The above formula is consistent with the usual Weyl quantisation on (a Poisson *-subalgebra of) the smaller classical algebra $C_0(T^*\T^n)$, see e.g. \cite[section II.3.4]{landsman98}, as $\T^n$ is in particular a Riemannian manifold with its corresponding Levi-Civita connection. Although this consistency already justifies \eqref{eq:formula QW intro} as a reasonable extension of Weyl quantisation, we start section \ref{sec:quantisation} with a systematic way to arrive at \eqref{eq:formula QW intro}.
Thereafter, we define the (quantum) resolvent algebra on the torus as
	$$A_\hbar:=C^*(\QW(\Sr(T^*\T^n)))\subseteq B(L^2(\T^n))\,,$$
before remarking that $A_\hbar\cong A_{\hbar'}$ for all $\hbar,\hbar'\in(0,\infty)$. We check \ref{property:direct} by using this explicit description of $\QW$ and the fact that \ref{property:direct} holds for $\Cr(T^*\T^n)$, which is readily seen. \ref{property:small} is satisfied by definition of $\Cr(T^*\T^n)$.
In addition, we show that an analogue of \ref{property:time} holds for our algebras, both the classical and the quantum one, in the following very strong sense: our classical resolvent algebra $C_{\mathcal{R}}(T^\ast \IT^n)$ is closed under the classical time evolution associated to the potential $V$ for each $V \in C^1(\IT^n)$ with Lipschitz continuous derivative.
Our quantum resolvent algebra is closed under the quantum time evolution associated to the potential $V$ for each $V \in C(\IT^n)$.
(In both cases, the free part of the Hamiltonian is the usual one.)
Unlike the analogous result in \cite{buchholz08} in which a similar result is established only for $\IR^{2n}$ with $n = 1$, we give proofs of these statements for arbitrary $n \in \IN$.\\

\noindent 
The paper is structured as follows.
In section 2, we first give a well-motivated definition of the classical resolvent algebra $C_{\mathcal{R}}(T^\ast \IT^n)$.
We proceed by analysing its structure, culminating in an alternative, more practical characterisation of $C_{\mathcal{R}}(T^\ast \IT^n)$, namely as the tensor product $C(\IT^n) \hatotimes \Teunalg{\IR^n}$.
Furthermore, we identify a dense Poisson $^\ast$-subalgebra that serves the same purpose as $\mathcal{S}_{\mathcal{R}}(\IR^n)$ in \cite{vnuland19}.

Section 3 proves the fact that $C_{\mathcal{R}}(T^\ast \IT^n)$ is closed under the classical time evolution as mentioned above.

In section 4, we adapt Weyl quantisation to functions on $T^\ast \IT^n$, proving an explicit formula for generators of $C_{\mathcal{R}}(T^\ast \IT^n)$ in the process.
This formula is then used to show that Weyl quantisation is almost a strict quantisation in the sense of Landsman.
We say `almost', because we explicitly show that its norm fails to be continuous with respect to $\hbar$ for $\hbar > 0$.
However, the quantisation map is continuous in a weaker sense.

In section 5, we show that our quantised resolvent algebra is closed under the quantum time evolution.

\noindent
\paragraph{Acknowledgements}
We would like to thank Francesca Arici, Victor Gayral, Klaas Landsman and Walter van Suijlekom for useful comments and helpful discussions.
This research was supported by the Dutch Research Council (NWO) under project numbers 639.031.827 and 680.91.101.

\ifx\mycmd\undefined
	\phantomsection
	\addcontentsline{toc}{section}{References}
	\bibliographystyle{abbrv}
	\bibliography{./../Miscellaneous/References}

	\end{document}
\fi

\ifx\mycmd\undefined
	\documentclass[a4paper, 11pt, leqno]{article}
	
	\begin{document}
\fi

\section{Definition and basic results}
\label{sct:Basic Results}

\noindent
On the phase space $\R^{2n}$, we already have a commutative C*-algebra that satisfies \ref{property:time}, \ref{property:direct} and \ref{property:small} mentioned in the introduction and forms the classical part of a strict deformation quantisation, namely the commutative resolvent algebra $\Cr(\R^{2n})$ defined in \cite{vnuland19}.
We begin this section by adapting its definition to $T^\ast \IT^n$.
As mentioned in the introduction, we identify $T^\ast \IT^n$ with $\IT^n \times \IR^n$, and note that the latter space carries a natural left action of $\IR^{2n} = \IR^n \times \IR^n$ by translation.

\begin{defi}
For each $(v,w) \in \IR^n \times \IR^n = \IR^{2n}$, let $(\IT^n \times \IR^n)/\{(v,w)\}^\perp$ be the space of orbits of the restriction of the action of $\IR^{2n}$ to $\{(v,w)\}^\perp \subseteq \IR^{2n}$, and let $$\pi_{(v,w)} \colon \IT^n \times \IR^n \rightarrow (\IT^n \times \IR^n)/\{(v,w)\}^\perp$$ be the corresponding canonical projection.
We then define the {\em commutative resolvent algebra $\Cr(T^\ast \IT^n)$} as the smallest C$^\ast$-subalgebra of $C_b(\T^n\times\R^n)$ generated by the set of functions
\begin{equation*}
\left\{ f \circ \pi_{(v,w)} \mid (v,w) \in \IR^{2n}, \: f \in C_0((\IT^n \times \IR^n)/\{(v,w)\}^\perp) \right\},
\end{equation*}
that is, the set of continuous functions invariant under the action of $\{(v,w)\}^\perp \subseteq \IR^{2n}$ for which the induced map on $(\IT^n \times \IR^n)/\{(v,w)\}^\perp$ vanishes at infinity.
\end{defi}

\noindent
To establish the link with the definition of $\Cr(\IR^n)$ given in \cite{vnuland19}, note that there is a straightforward generalisation of the above definition to arbitrary topological spaces $M$ carrying a left action of $\IR^{m}$ for some $m \in \IN$.
Taking $M = \IR^n$ and $m = n$ then yields the definition of $\Cr(\IR^n)$.
Unfortunately, $T^*G$ does not have an appropriate action of $\R^{2n}$ for a nonabelian Lie group $G$ that would enable us to unambiguously generalise this construction.

The definition of the classical resolvent algebra $C_{\mathcal{R}}(T^\ast \IT^n)$ is clearly motivated, but very unwieldy in practice.
Our first task is therefore to find an alternative, more elementary characterisation of $C_{\mathcal{R}}(T^\ast \IT^n)$.
To this end we will use the following elementary facts about the action of $\R^n$ on $\T^n$. Throughout the rest of the text, we let $[x]\in\T^n=\R^n/\Z^n$ denote the quotient class of $x\in\R^n$.

\begin{lem}\label{lem: dense or periodic}
Let $v \in \R^n \backslash \{0\}$.
\begin{myenum}
\item Exactly one of the following two statements holds true:
\begin{enumerate}[(i)]
\item The map $\R\rightarrow\T^n$, $t\mapsto[tv]$ is periodic.
\item The set $H:=\set{[x]\in\T^n}{ x\in\R^n,\, v\cdot x=0 }$ is dense in $\T^n$.
\end{enumerate}
\item Suppose now that $t\mapsto[tv]$ is periodic, with period $T$.
Furthermore, let $\pi_v \colon \IT^n \rightarrow \IT^n/H$ be the quotient map.
Then $H$ is a closed subgroup of $\T^n$, and
\begin{equation*}
\varphi \colon \T^n/H \rightarrow \T, \quad 
\pi_v([x]) \mapsto \left[ Tv\cdot x \right]\,,
\end{equation*}
is a well-defined isomorphism of topological groups.
\end{myenum}
\end{lem}

\begin{proofenum}
\item We show that at least one of the two statements is true; we postpone the proof that the two statements are mutually exclusive to the proof of the second part of this lemma.
The case $n = 1$ is trivial, and the case $n = 2$ is the well-known result that a line in $\mathbb{T}^2$ is dense iff it has irrational slope.
	We therefore assume that $n>2$, and we will reduce the problem to the known two-dimensional case.

	Suppose that $(ii)$ is false for $n>2$, and let $U\subseteq \T^n\backslash H$ be a non-empty open subset. Without loss of generality we may assume that $U=[y]+U$ for all $y\perp v$. As $v\neq0$ by assumption, we may choose $j$ such that $v_j\neq0$. Now suppose $k$ is a different index with $v_k\neq0$.
	Define
$$T_2:=\set{[x]\in\T^n}{ x\in\R^n,\,x_i=0 \textnormal{ for all }i\notin\{j,k\}}\subseteq\T^n\,,$$
	which is a subgroup of $\T^n$ isomorphic to $\T^2$.
	Note that 
		$$U\cap T_2\subseteq T_2\setminus\set{[x]\in T_2}{ x\in\R^n,\,v_jx_j+v_kx_k=0}\,.$$
	To see that $U\cap T_2$ is non-empty, let us pick $[x]\in U$.
	Because $v_j\neq0$, there exists a $y \perp v$ such that $[y+x]\in T_2$ (for instance $y=\frac{v\cdot x}{v_j}\delta_j-x$, where $\delta_j$ is the $j^{\text{th}}$ standard basis vector). 
	Since $[y+x]\in [y]+U=U$ we have $[y+x]\in U\cap T_2\neq \emptyset$.
	Applying the result for $n=2$, one finds that $[t(v_j,v_k)]$ is periodic in $t$. Since $k$ (such that $v_k\neq0$) was arbitrary, every component $v_k$ is a rational multiple of the nonzero component $v_j$, hence $[tv]$ is periodic in $t$.

\item We first note that the map $\varphi$ is induced (in two steps) by the continuous group homomorphism
	\begin{equation*}
		\R^n \rightarrow \T \, , \quad 
		x \mapsto \left[ T v \cdot x \right]\,.
	\end{equation*}
	Since $Tv\in \Z^n$, this map factors through $\T^n$, thereby inducing a continuous group homomorphism $\varphi_0 \colon \IT^n \rightarrow \IT$.
	It is readily seen that $H$ is a subgroup of $\T^n$ that is a subset of $\ker \varphi_0$, so $\varphi_0$ factors through the quotient $\T^n / H$, thereby inducing the continuous group homomorphism $\varphi$.

	Next, we argue that $\varphi$ is in fact a homeomorphism.
	We prove this by showing that the map
	\begin{equation*}
		\T \rightarrow \T^n / H \, , \quad 
		[t] \mapsto \pi_v\left(\left[\frac{tv}{T\norm{v}^2}\right]\right) \, ,
	\end{equation*}
	is a well-defined inverse; we will tentatively refer to this map as $\varphi^{-1}$ in what follows.

	First we show that $\varphi^{-1}$ is well-defined.
	This amounts to showing that
	\begin{equation*}
		\frac{v}{T\norm{v}^2} \cdot \IZ \subseteq \set{a + x}{a \in \IZ^n, \: x \in \{v\}^\perp} \, ,
	\end{equation*}
	which is the case exactly when $v/(T\norm{v}^2)$ is an element of the set on the right-hand side of the inclusion.
	Note that the components of the vector $Tv$ are coprime; otherwise, $T/m$ would be the period of $t\mapsto[tv]$ for some natural number $m > 1$.
	By the higher-dimensional B\'ezout identity, there exists a tuple $a \in \Z^n$ such that $Tv\cdot a = 1$.
	Now observe that
	\begin{equation*}
		\frac{v}{T\norm{v}^2}
		= \frac{v \cdot a}{\|v\|^2}v
		=  a + \left(\frac{v \cdot a}{\|v\|^2}v - a\right) \, ,
	\end{equation*}
	and note that the first and second term on the right-hand side of this equation are contained in $\Z^n$ and $\{v\}^\perp$, respectively.
	Thus $\varphi^{-1}$ is well-defined.
	It is straightforward to check that $\varphi^{-1}$ is both a left- and a right-inverse of $\varphi$, so $\varphi^{-1}$ is indeed the inverse of $\varphi$.
	It is readily seen that $\varphi^{-1}$ is continuous, so $\varphi$ is both a group isomorphism and a homeomorphism.

	To see that $\varphi$ is in fact an isomorphism of topological groups, note that $H = \ker \varphi_0$ by injectivity of $\varphi$, so $H$ is a closed subgroup of $\T^n$, and the quotient $\T^n / H$ naturally inherits the structure of a topological group from $\T^n$; in particular, the quotient is Hausdorff.
This concludes the proof of part (2).

Finishing up the proof of part (1), we note that in case (i), the quotient $\IT^n / H$ is homeomorphic to $\IT$, whereas in case (ii), the quotient is an indiscrete space.
Thus the two cases are mutually exclusive.
\end{proofenum}

\begin{rem}
In case (i), $H$ is a Lie subgroup of the Lie group $\T^n$ by the closed-subgroup theorem.
In fact, $\varphi$ is an isomorphism of Lie groups from $\IT^n / H$ to $\T$ endowed with their respective canonical Lie group structures.
See \cite[Lemma 5.4]{stienstra19} for a Lie-theoretic version of the previous lemma.
\end{rem}

\noindent 
Before we characterise $\Cr(T^\ast \IT^n)$, we must introduce another algebra, for which it is in turn useful to recall that the algebra of {\em almost periodic functions on $\IR^n$} is the C$^\ast$-subalgebra of $C_b(\IR^n)$ generated by functions of the form $x \mapsto e^{i \xi \cdot x}$, where $\xi \in \IR^n$ is arbitrary.
Almost periodic functions were originally introduced by H. Bohr in \cite{bohr25I} for $n = 1$ using a different definition, whose equivalence with the one mentioned above he proved in \cite{bohr25II}.
This algebra will be denoted by $\Weylalg{\IR^n}$.

\begin{defi}
Let $n \in \IN$.
We define the algebra $\Teunalg{\IR^n}$ as the C$^\ast$-subalgebra of $C_b(\IR^n)$ generated by the classical resolvent algebra $C_{\mathcal{R}}(\IR^n)$ and the algebra of almost periodic functions $\Weylalg{\IR^n}$ on $\IR^n$.
\end{defi}

\noindent 
Next up is the main result of this section, which unveils $\Cr(T^*\T^n)$ as a tensor product of two algebras. We regard the algebraic tensor product of two C*-algebras $A\subseteq C_b(X)$ and $B\subseteq C_b(Y)$ as a subset of $C_b(X\times Y)$ via $(f\otimes g)(x,y)=f(x)g(y)$, and denote its corresponding completion by $A\hatotimes B$.
Since commutative C$^\ast$-algebras are nuclear (cf. \cite[Theorem 6.4.15]{murphy90}), this is equivalent to any other C*-algebraic tensor product.

\begin{thrm}\label{thrm:tractable_resolvent_algebra}
For each $n \in \IN$, we have
\begin{equation*}
C_{\mathcal{R}}(T^\ast \IT^n) = C(\IT^n) \hatotimes \Teunalg{\IR^n}\,.
\end{equation*}
\end{thrm}
\begin{proof}
The statement is trivial for $n = 0$, so suppose $n \geq 1$.
We first prove the inclusion $C_{\mathcal{R}}(T^\ast \IT^n) \subseteq C(\IT^n) \hatotimes \Teunalg{\IR^n}$ by showing that the generators of $C_{\mathcal{R}}(T^\ast \IT^n)$ are contained in the right-hand side.
Let $(v,w) \in \IR^{n}\times\R^n$, and let $f=g\circ \pi_{(v,w)}$ be one of the generators of $C_{\mathcal{R}}(T^\ast \IT^n)$.
As in Lemma \ref{lem: dense or periodic}, let $H$ be the image of $\{v\}^\perp$ under the canonical projection map $\IR^n \rightarrow \IT^n$.
Moreover, let $H^\prime$ be the image of $\{(v,w)\}^\perp$ under the canonical projection map $\IR^n \times \IR^n \rightarrow \IT^n \times \IR^n$.

By part (1) of Lemma \ref{lem: dense or periodic} we may distinguish between the following three cases.
In each of these cases, we obtain the general form of $f$ by first giving a characterisation of the quotient space $(\IT^n \times \IR^n) / H^\prime$:
\begin{enumerate}[label={(\roman*) \:}, wide, labelwidth=!, labelindent=0pt, parsep=0pt]
\item $v = 0$: in this case, we have $H^\prime = \IT^n \times \{w\}^\perp$; in particular, it is a closed subgroup of $\IT^n \times \IR^n$, and the map
\begin{equation*}
(\IT^n \times \IR^n)/H^\prime \rightarrow \IR \cdot w \, , \quad 
\pi_{(v,w)}([x], p) \mapsto (w \cdot p)w \, ,
\end{equation*}
is an isomorphism of topological groups.
It follows that $f$ is the pullback of a function in $C_0(\IR \cdot w)$ along the above map, from which it is readily seen that
\begin{equation*}
f \in \IC \unit_{\IT^n} \hatotimes C_{\mathcal{R}} (\IR^n)
\subseteq C(\IT^n) \hatotimes \Teunalg{\IR^n} \,;
\end{equation*}
In particular, note that $f$ is constant iff $w = 0$.
\end{enumerate}
To handle the remaining two cases in which $v \neq 0$, we introduce the map
\begin{align*}
\theta \colon (\IT^n \times \IR^n) / H^\prime &\rightarrow \IT^n / H \, , \\
\pi_{(v,w)}([x],p) &\mapsto \pi_v \left( \left[ \frac{v \cdot x + w \cdot p}{\|v\|^2} v \right] \right)
= \pi_v \left( \left[ x + \frac{w \cdot p}{\|v\|^2} v \right] \right) \, ,
\end{align*}
and show that it is a well-defined group isomorphism and a homeomorphism.
To see that it is a well-defined continuous group homomorphism, note that it is induced by a continuous group homomorphism
\begin{equation*}
\theta_0 \colon \IT^n \times \IR^n \rightarrow \IT^n / H \, ,
\end{equation*}
which is defined using a similar formula, and whose kernel contains the subgroup $H^\prime$.
To see that $\theta$ is a group isomorphism and a homeomorphism, we note that the map
\begin{equation*}
\IT^n / H \rightarrow (\IT^n \times \IR^n) / H^\prime \, , \quad 
\pi_v ([x]) \mapsto \pi_{(v,w)}([x],0) \, ,
\end{equation*}
is a well-defined countinuous group homomorphism (by a similar argument as for $\theta$) that can be checked to be the inverse of $\theta$.
In particular, $H^\prime = \ker \theta_0$.
As we will see below, $\theta$ need not be an isomorphism of topological groups if we require such groups to be Hausdorff spaces.
We proceed with the remaining two cases:
\begin{enumerate}[label={(\roman*) \:}, wide, labelwidth=!, labelindent=0pt, parsep=0pt]
\setcounter{enumi}{1}
\item $v \neq 0$ and $H$ is dense in $\IT^n$: in this case, the quotient topology on $\IT^n / H$ is the indiscrete topology, hence $(\IT^n \times \IR^n) / H^\prime$ is also indiscrete by our discussion above.
It follows that the function $f$ is constant, so $f \in C(\IT^n) \hatotimes \Teunalg{\IR^n}$.

\item $v \neq 0$ and the curve $t \mapsto [tv]$ on $\IT^n$ is periodic: then the map $\theta_0$ defined above is a continuous surjective group homomorphism, hence its kernel $H^\prime$ is a closed subgroup of $\IT^n \times \IR^n$, and the map $\theta$ is an isomorphism of topological groups.
Composing $\theta$ with the map $\varphi$ from part (2) of Lemma \ref{lem: dense or periodic}, we obtain the isomorphism of topological groups
\begin{equation*}
\varphi \circ \theta \colon (\IT^n \times \IR^n)/H^\prime \rightarrow \IT \, , \quad 
\pi_{(v,w)}([x], p) \mapsto \left[ T(v \cdot x + w \cdot p) \right] \, ,
\end{equation*}
with $T$ as defined in Lemma \ref{lem: dense or periodic}.
Then $f = g \circ \varphi \circ \theta \circ \pi_{(v,w)}$ for some $g \in C(\IT)$; let us first assume that $g=e_k$ for some $k \in \IZ$.
Then
\begin{align*}
f([x],p)
&= \exp \left( 2\pi i kT(v \cdot x + w \cdot p) \right) \\
&= \exp \left( 2\pi i kTv \cdot x \right) \cdot \exp \left( 2\pi i kT w \cdot p \right) \, ,
\end{align*}
which shows that $f \in C(\IT^n) \hatotimes \Teunalg{\IR^n}$.
Since the family of exponential functions $(e_k)_{k \in \Z}$ generate $C(\IT)$, and since pullback along the map
\begin{equation*}
\varphi \circ \theta \circ \pi_{(v,w)} \colon \IT^n \times \IR^n \rightarrow \IT \, ,
\end{equation*}
is a homomorphism of C$^\ast$-algebras, it follows that
\begin{equation*}
f = g \circ \varphi \circ \theta \circ \pi_{(v,w)} \in C(\IT^n) \hatotimes \Teunalg{\IR^n} \, ,
\end{equation*}
for arbitrary $g \in C(\IT)$.
\end{enumerate}
This establishes the inclusion $C_{\mathcal{R}}(T^\ast \IT^n) \subseteq C(\IT^n) \hatotimes \Teunalg{\IR^n}$.
The reverse inclusion is a consequence of the following three observations:
\begin{itemize}
\item \sloppy From case (i) in the previous part of this proof, we readily obtain $\IC \unit_{\IT^n} \hatotimes C_{\mathcal{R}}(\IR^n) {{} \subseteq C_{\mathcal{R}}(T^\ast \IT^n)}$.

\item From case (iii), setting $w = 0$ and taking $v$ to be a standard basis vector of $\IR^n$, we obtain $C(\IT^n) \hatotimes \IC \unit_{\IR^n} \subseteq C_{\mathcal{R}}(T^\ast \IT^n)$.

\item Finally, by considering case (iii) again, but now with $v$ the first standard basis vector and $w \in \IR^n$ arbitrary, we see that $C_{\mathcal{R}}(T^\ast \IT^n)$ contains functions of the form
\begin{equation*}
([x], p) \mapsto \exp(2 \pi i k x_1) \exp(i \xi \cdot p) \, ,
\end{equation*}
where $k \in \IZ \backslash \{0\}$, and $\xi \in \IR^n$ is arbitrary.
The previous point now implies that functions of the form
\begin{equation*}
([x], p) \mapsto \exp(i \xi \cdot p) \, ,
\end{equation*}
are elements of the resolvent algebra, so $\IC \unit_{\IT^n} \hatotimes \Weylalg{\IR^n} \subseteq C_{\mathcal{R}}(T^\ast \IT^n)$. \qedhere
\end{itemize}
\end{proof}

\noindent 
We finish this section by defining the analogue of the space of Schwartz functions of $C_{\mathcal{R}}(T^\ast \IT^n)$.
This allows us to introduce the notation $h_{U,\xi,g}$ for the generators of $\Teunalg{\IR^n}$, which will be used in section \ref{sec:quantisation}.

\begin{defi}
For each $k \in \IZ^n$, let
\begin{equation*}
e_k \colon \IT^n \rightarrow \IC, \quad 
[x] \mapsto e^{2\pi i k \cdot x}\,.
\end{equation*}
For each subspace $U \subseteq \IR^n$, for each $\xi \in U^\perp$, and for each Schwartz function $g \in \mathcal{S}(U)$, let
\begin{equation*}
h_{U,\xi,g} \colon \IR^n \rightarrow \IC, \quad 
p \mapsto e^{i\xi \cdot p} g ( P_U(p))\,,
\end{equation*}
where $P_U \colon \IR^n \rightarrow U$ denotes the orthogonal projection onto $U$.
We define the space $\mathcal{S}_{\mathcal{R}}(T^\ast \IT^n)$ as the span of functions of the form $e_k\otimes h_{U,\xi,g}:\T^n\times\R^n\rightarrow\C$.
\label{def:resolvent_algebra_Schwartz_functions}
\end{defi}

\begin{prop}

\noindent
\begin{myenum}
\item The space $\spn\set{h_{U,\xi,g}}{ U\subseteq \R^n \text{ linear},~\xi\in U^\perp,~g\in\S(U)}$ is a norm-dense $^\ast$-subalgebra of $\Teunalg{\IR^n}$ that is closed under multiplication and partial differentiation.

\item The vector space $\mathcal{S}_{\mathcal{R}}(T^\ast \IT^n)$ is a subspace of $C_{\mathcal{R}}(T^\ast \IT^n)$ that is closed under multiplication and partial differentiation, and is consequently a Poisson subalgebra of $C^\infty(T^\ast \IT^n)$.
Moreover, $\mathcal{S}_{\mathcal{R}}(T^\ast \IT^n)$ is a norm-dense $^\ast$-subalgebra of $C_{\mathcal{R}}(T^\ast \IT^n)$.
\end{myenum}
\label{prop:Poisson_subalg_for_arbitrary_n}
\end{prop}

\begin{proofenum}
\item Denote $\mB:=\spn\{h_{U,\xi,g}\}\subset\Wr$. For any $h_{U,\xi,g}$ as in Definition \ref{def:resolvent_algebra_Schwartz_functions}, 
\begin{equation*}
h_{U,\xi,g}^\ast
= \overline{h_{U,\xi,g}}
= h_{U,-\xi,\overline{g}}
\in \mathcal{B}\,,
\end{equation*}
hence $\mathcal{B}$ is closed under the $^\ast$-operation.

Assume for the moment that $\mathcal{B}$ is closed under multiplication.
To see that $\mathcal{B}$ is invariant under partial differentiation, it suffices to show that partial derivatives of functions of the form $h_{U,\xi,g}$ are elements of $\mathcal{B}$.
Any partial derivative can be written as a sum of two directional derivatives; one in a direction lying in $U$, and one in a direction lying in $U^\perp$.
It is readily seen that both of these directional derivatives are elements of $\mathcal{B}$, hence so is their sum.

To show that $\mathcal{B}$ is closed under multiplication, it suffices to show that the product of two functions $h_{U_1,\xi_1,g_1}$ and $h_{U_2,\xi_2,g_2}$ as in Definition \ref{def:resolvent_algebra_Schwartz_functions}, is an element of $\mathcal{B}$.
Let 
\begin{align*}
U &:= U_1 + U_2\,, \\
\xi &:= \xi_1 + \xi_2 - P_U(\xi_1 + \xi_2) \in U^\perp\,, \\
\tilde{g} &:= (g_1 \circ P_{U_1})(g_2 \circ P_{U_2})\,.
\end{align*}
Note that the restrictions of $\tilde{g}$ to $U$ and $U^\perp$ are Schwartz and constant, respectively.
Setting
\begin{equation*}
g \colon U \rightarrow \IC\,, \quad 
p \mapsto e^{i P_U(\xi_1 + \xi_2) \cdot p} \tilde{g}|_U \circ P_U(p)
= e^{i (\xi_1 + \xi_2) \cdot p} \tilde{g}|_U \circ P_U(p)\,,
\end{equation*}
we see that $h_{U_1,\xi_1,g_1} \cdot h_{U_2,\xi_2,g_2} = h_{U,\xi,g}$, which establishes that $\mathcal{B}$ is closed under multiplication.

Thus $\mathcal{B}$ is a $^\ast$-subalgebra of $\Teunalg{\IR^n}$.
In addition to this fact, the elements of the form $h_{\{0\},\xi,1}$ generate $\Weylalg{\IR^n}$, while the elements of the form $h_{U,0,g}$ generate $C_{\mathcal{R}}(\IR^n)$, hence $\mathcal{B}$ generates $\Teunalg{\IR^n}$ as a C$^\ast$-algebra.
We infer that $\Teunalg{\IR^n}$ is the closure of $\mathcal{B}$.

\item For each $k \in \IZ^n$, define $e_k$ as in Definition \ref{def:resolvent_algebra_Schwartz_functions}.
It is a trivial matter to check that the linear span of $\set{e_k}{k \in \IZ}$ is a $^\ast$-subalgebra of $C(\IT^n)$ that is closed with respect to partial differentiation, and it is a result from Fourier analysis that this linear subspace is dense in $C(\IT^n)$.
Using these facts in conjunction with part (1) of this proposition and Theorem \ref{thrm:tractable_resolvent_algebra}, it is readily seen that all of the assertions are true.
\end{proofenum}

% Rieffel: action of $\IR^n$ on $\IR^n$.

% Disclaimer: only $\IT^n$ instead of arbitrary compact Lie groups, and not defined in terms of generators and relations.
\ifx\mycmd\undefined
	\phantomsection
	\addcontentsline{toc}{section}{References}
	\bibliographystyle{abbrv}
	\bibliography{./../Miscellaneous/References}

	\end{document}
\fi

\ifx\mycmd\undefined
	\documentclass[a4paper, 11pt, leqno]{article}
	
	\begin{document}
\fi

\section{Classical time evolution}\label{sec:classical time evolution}
%
%\paragraph{Notation.}
%In contrast with the other sections, we use the letter $q$ for points in $\T^n$, thus avoiding the notation $[q]$.  \\

\noindent In this section, we prove that $\Cr(T^*\T^n)$ is preserved under the (time) flow induced by the Hamiltonian
	$$H(q,p)=\tfrac{1}{2}p^2 + V(q)\,,$$
for each potential $V\in C^1(\T^n)_\R$ such that $\nabla V$ is Lipschitz continuous.
This is arguably the most natural assumption on $V$; the Picard--Lindel\"of theorem then ensures that the Hamilton equations have unique solutions. 

Precisely stated, for every $(q_0,p_0)\in \T^n\times\R^n$, there exist unique functions $q \colon \R \rightarrow \T^n$ and $p \colon \R \rightarrow \R^n$ that satisfy
\begin{equation}
\label{eq:q and p}
\left\{
\begin{alignedat}{2}
(\dot{q}(t),\dot{p}(t)) &= (p(t),-\nabla V(q(t))) \qquad && t \in \R\,, \\
(q(0),p(0)) &= (q_0, p_0)\,. \qquad &&
\end{alignedat}
\right. 
\end{equation}
Note that the expression on the right-hand side of the first line of equation \eqref{eq:q and p} is the Hamiltonian vector field $X_H$ corresponding to $H$ evaluated at $(q(t),p(t))$.
For each $t \in \R$, the time evolution of the system after time $t$ is the map
\begin{equation*}
\Phi^t_V \colon \T^n\times\R^n \rightarrow \T^n\times\R^n, \quad 
(q_0,p_0)\mapsto(q(t),p(t)),
\end{equation*}
which is the flow corresponding to $X_H$ evaluated at time $t$; it is well-known to be a homeomorphism.

Note that we have already made the notation of the flow less cumbersome by writing $\Phi^t_V$ instead of $\Phi^t_{X_H}$.
In what follows, we restrict our attention to the case $t = 1$, further simplifying the notation by defining $\Phi_V:=\Phi^1_V$.
The following lemma shows that we may do so without loss of generality:

\begin{lem}\label{lem:t=1 or arbitrary t}
	The algebra $\Cr(T^*\T^n)$ is preserved under the pullback of $\Phi_V$ for each $V$ if and only if it is preserved under the pullback of $\Phi_V^t$ for each $V$, for each $t\in\R$.
\end{lem}
\begin{proof}
	For any $t\neq0$ (as $t=0$ is trivial), we make the following transformation on phase space
	$$\phi(q,p):=(q,tp)\,.$$
	Because the momentum part of $\phi$ is linear, its pullback preserves the commutative resolvent algebra. Given an integral curve $(q(t),p(t))$ of the vector field $X_H$ corresponding to the potential $V$, i.e., a solution of equation \eqref{eq:q and p}, one can easily check that $s\mapsto\phi(q(ts),p(ts))$ is an integral curve corresponding to the potential $t^2V$. We therefore conclude that
	$$\Phi_V^t(q_0,p_0)=\phi^{-1} \circ \Phi^1_{t^2V} \circ \phi(q_0,p_0)\,,$$
	which implies the claim.
\end{proof}
\noindent We prove our main theorem in three steps: taking $V=0$; taking $V$ trigonometric; and finally taking general $V$.
In the second and third step we will need the following consequence of Gronwall's inequality.
Let $d$ denote the canonical distance function on $\T^n$ as well as on $\T^n\times\R^n$. (Note that these distance functions are the ones induced by the canonical Riemannian metrics on $\T^n$ and $T^*\T^n\cong\T^n\times\R^n$, respectively.)
\begin{lem}\label{lem:Gronwall}
	Let $f,g\colon\T^n\times\R^n\rightarrow\R^{2n}$ be Lipschitz continuous functions, let $c$ be the Lipschitz constant of $f$, and let $y,z\colon[0,1]\rightarrow \T^n\times\R^n$ be curves that satisfy $\dot{y}(t)=f(y(t))$ and $\dot{z}(t)=g(z(t))$ for each $t\in[0,1]$.
Finally, suppose that $\varepsilon > 0$ is a number such that $\supnorm{f-g}\leq\varepsilon$.
Then we have
	$$d(y(t),z(t))\leq(d(y(0),z(0))+t\varepsilon)e^{tc}\,.$$
\end{lem}
\begin{proof}
	By translation invariance of the metric on $\T^n\times\R^n$, we have
	\begin{align*}
		d(y(t),z(t))&\leq d((y(t)-y(0))-(z(t)-z(0)),0)+d(y(0),z(0))\\
		&\leq\int_0^t\norm{f(y(s))-g(z(s))}\:ds+d(y(0),z(0))\\
		&\leq c\int_0^t d(y(s),z(s))\:ds+t\varepsilon+d(y(0),z(0))\,.
	\end{align*}
	With the integral version of Gronwall's inequality, this implies the lemma.
\end{proof}

\subsection{Free time evolution}\label{sct:free time evolution}

\noindent
For each pair $(q_0,p_0)\in \T^n\times\R^n$, we have $q(t)=q_0+tp_0$ and $p(t)=p_0$, denoting the usual action of $\R^n$ on $\T^n$ by $+$.
%For each pair $(q_0,p_0)\in \T^n\times\R^n$, we have $q(t)=q_0+tp_0$ and $p(t)=p_0$.
%Here, we have made an abuse of notation by omitting the square brackets around $tp_0$, thereby obfuscating the distinction between elements of $\IT^n$ and $\IR^n$.
The latter notation, explicitly written as $[x]+p=[x+p]$ for $x,p\in\R^n$, will be used in the remainder of this section.
We find that $\Phi_0(q_0,p_0)=(q_0+p_0,p_0)$, and obtain the following preliminary result.
Let ${}^*$ denote the pullback.
\begin{lem}\label{lem:free time evolution}
	Free time evolution preserves the commutative resolvent algebra, i.e.,
	$$\Phi_0^*(\Cr(T^*\T^n))\subseteq\Cr(T^*\T^n)\,.$$
\end{lem}
\begin{proof}
We have
	$$\Phi_0^*(e_k\otimes h_{U,\xi,g})(q_0,p_0)=e_k(q_0)e^{ 2\pi i k\cdot p_0}e^{i\xi\cdot p_0}g(P_U (p_0))\,.$$
Defining $\tilde{g}\in C_0(U)$ by $\tilde{g}(p):=e^{2\pi iP_U(k)\cdot p}g(p)$, and $\tilde{\xi}:=\xi+2\pi P_{U^\perp}(k)$, we obtain
	$$\Phi_0^*(e_k\otimes h_{U,\xi,g})=e_k\otimes h_{U,\tilde{\xi},\tilde{g}}\,.$$
Thus the generators of $\Cr(T^*\T^n)$ are mapped into $\Cr(T^*\T^n)$ by $\Phi_0^*$, and since this map is a $^\ast$-homomorphism, the lemma follows.
\end{proof}

\subsection{Trigonometric potentials}\label{sct:trigonometric potentials}

We say that $V$ is a {\em trigonometric potential} if it is real-valued and of the form $V=\sum_{k\in\mN}a_ke_k$, for some coefficients $a_k\in\C$ and a finite subset $\mN\subseteq\Z^n$. The main trick used to establish time invariance of the classical resolvent algebra is to use induction on the size of $\mN$. The induction basis, $\mN=\emptyset$, corresponds to free time evolution. In order to carry out the induction step we fix a vector $k\in\mN$, and compare the dynamics corresponding to $V$ with the dynamics corresponding to $V-V_k$, where $$V_k:=a_ke_k+a_{-k}e_{-k}\,.$$
Similar to the already defined curves $q \colon [0,1]\rightarrow\T^n$ and $p \colon [0,1]\rightarrow\R^n$, the dynamics corresponding to $V-V_k$ of the point $(q_0,p_0)$ is incapsulated by the curves $\tilde{q} \colon[0,1]\rightarrow\T^n$ and $\tilde{p} \colon [0,1]\rightarrow\R^n$ satisfying
\begin{equation}
\label{eq:tilde_q and tilde_p}
\left\{
\begin{alignedat}{2}
(\dot{\tilde{q}}(t),\dot{\tilde{p}}(t)) &= (\tilde{p}(t),-\nabla (V - V_k)(\tilde{q}(t))) \qquad && t \in \R\,, \\
(\tilde{q}(0),\tilde{p}(0)) &= (q_0, p_0)\,. \qquad &&
\end{alignedat}
\right. 
\end{equation}
We compare the two dynamics in the following proposition.

\begin{prop}\label{lem:long range convergence of dynamics}
	Let $k\in\Z^n$ and $\delta>0$. There exists a $D_k>0$ such that for each $(q_0,p_0)\in \T^n\times\R^n$ satisfying $|k\cdot p_0|>D_k$, we have
		$$ d\left(\Phi_V(q_0,p_0),\Phi_{V-V_k}(q_0,p_0)\right)<\delta\,.$$
\end{prop}
\begin{proof}
	Note that the statement is vacuously true for any $D_k > 0$ if $k = 0$.
We therefore fix a nonzero $k\in\Z^n$. Throughout the proof, we use a variation of big O notation, expanding in the variable $\dt:=|k\cdot p_0|^{-1}$, uniformly in $q_0$. That is, we write $f(q_0,p_0)=\O(\dt^d)$ if there exist $N,C>0$ such that for all $q_0,p_0$ with $|k\cdot p_0|>N$ we have $|f(q_0,p_0)|\leq C|k\cdot p_0|^{-d}$.
	Therefore, to prove the proposition, it suffices to show that
	\begin{align}\label{suffice to show big O}
		d\left(\vect{q(1)}{p(1)},\vect{\tilde{q}(1)}{\tilde{p}(1)}\right) = \O(\dt)\,.
	\end{align}
	Assume that $\dt\in(0,1)$. We divide the time interval $[0,1]$ into $m$ intervals of length $\dt$, where $m:=\lfloor \frac{1}{\dt}\rfloor$, and a final interval of length $1-m\dt$.
For each $t\in[0,\dt]$ and each $j\in\{0,\ldots, m\}$ (these will be the assumptions on $t$ and $j$ throughout the rest of the proof), let
\begin{equation*}
q^j(t) := q(j\dt + t)\,, \quad p^j(t) := p(j\dt + t)\,,
\end{equation*}
and define the curves $\tilde{q}^j$ and $\tilde{p}^j$ analogously.
Note that $(q^j,p^j)$ and $(\tilde{q}^j,\tilde{p}^j)$ satisfy the differential equations \eqref{eq:q and p} and \eqref{eq:tilde_q and tilde_p} respectively, but with different initial conditions.
Furthermore, for every $j$, we define the curve $\gamma^j \colon [0,\dt] \rightarrow \T^n$ as the unique solution to the initial value problem
\begin{equation}
\label{eq:the_middle_man_gamma}
\left\{
\begin{alignedat}{2}
(\dot{\gamma}^j(t),\ddot{\gamma}^j(t)) &= (\dot{\gamma}^j(t),-\nabla (V - V_k)(\gamma^j(t))) \qquad && t \in \R\,, \\
(\gamma^j(0),\dot{\gamma}^j(0)) &= (q^j(0), p^j(0))\,. \qquad &&
\end{alignedat}
\right. 
\end{equation}
where on the first line, we have emphasised the similarity of this equation with the equations \eqref{eq:q and p} and \eqref{eq:tilde_q and tilde_p} by including $\dot{\gamma}^j(t)$.
We do not introduce any special notation for $\dot{\gamma}^j$, however.

\begin{figure}[!htb]
	\centering
	\hspace{0.02cm}
	\includegraphics[scale=0.125]{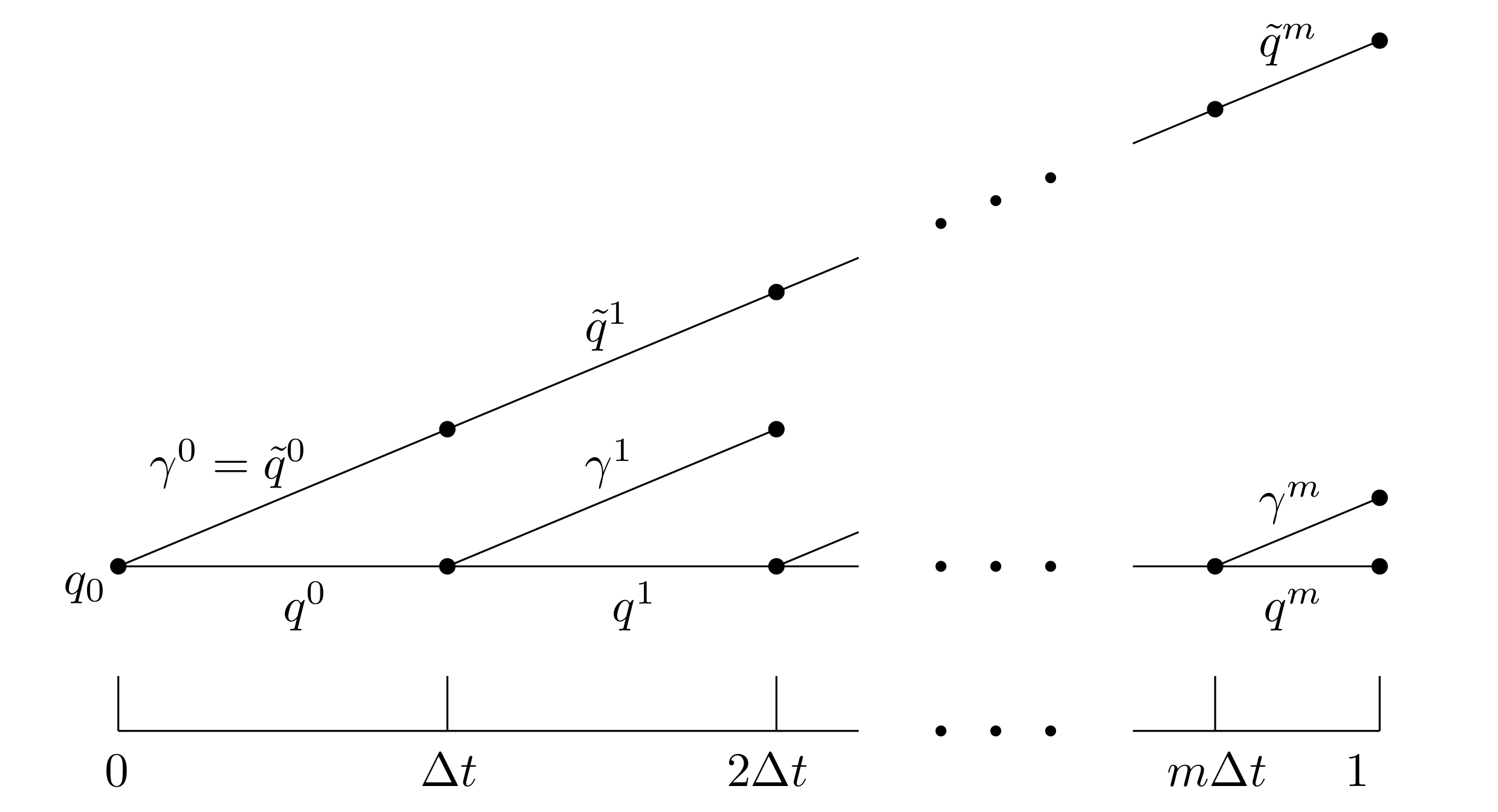}
	\vspace{-0.18cm}
	\caption{The position functions $q^j$,$\gamma^j$ and $\tilde{q}^j$. Sloping lines correspond to $V-V_k$, whereas the horizontal line that depicts $q$ corresponds to $V$.\label{fig: dynamica}}
\end{figure}	
	\noindent As depicted in Figure \ref{fig: dynamica}, the curve $\gamma^j\colon[0,\dt]\rightarrow\T^n$ plays a key r\^ole in 
comparing $q^j$ with $\tilde{q}^j$; the curve $(\gamma^j, \dot{\gamma}^j)$ is an integral curve along the same Hamiltonian vector field as $(\tilde{q}^j, \tilde{p}^j)$, but with the same initial conditions as $(q^j,p^j)$.
	
	We now expand our expressions in orders of $\dt$. Using equation \eqref{eq:q and p} and the fundamental theorem of calculus, we obtain
	\begin{align}\label{p_j}
		\norm{p^j(t)-p^j(0)}&\leq \int_0^{\dt}\norm{\nabla V (q^j(s))}\:ds\leq \supnorm{\nabla V}\dt
		=\O(\dt)\,.
	\end{align}
	In particular, taking $t=\dt$, we get $\norm{p^{j+1}(0)-p^j(0)}=\O(\dt)$, and therefore by induction
	\begin{align}\label{pj(0)-p0}
		\norm{p^j(0)-p_0}=\O(1)\,,
	\end{align}
	for every $0\leq j\leq m$. Equations \eqref{p_j} and \eqref{pj(0)-p0} give us
	\begin{align}\label{q^j}
		d(q^j(t),q^j(0)+tp_0) &\leq\norm{\int_0^{t}(p^j(s)-p_0)\:ds}\nonumber\\
		&\leq\int_0^{\dt}\norm{p^j(s)-p^j(0)}+\norm{p^j(0)-p_0}\:ds\nonumber\\
		&=\O(\dt)\,.
	\end{align}
	A result similar to \eqref{p_j} exists for $\dot{\gamma}^j$ instead of $p^j$, and hence
	\begin{align}\label{p and Gamma dot}
		\norm{p^j(t)-\dot{\gamma}^j(t)}=\O(\dt)\,,
	\end{align}
	which implies
	\begin{align}
		d(q^j(t),\gamma^j(t))&=\O(\dt^2)\,.\label{q^j and Gamma^j}
	\end{align}
	Using the definitions of $V_k$ and $\dt$, we show that the distance between $p^j(\dt)$ and $\dot{\gamma}^j(\dt)$ is in fact of order $\dt^2$. We first note that
	\begin{align*}
		\norm{p^j(\dt)-\dot{\gamma}^j(\dt)}&=\norm{\int_0^{\dt}(\nabla V(q^j(s))-\nabla(V-V_k)(\gamma^j(s)))\:ds}\\
		&\leq\int_0^{\dt}\norm{\nabla(V-V_k)(q^j(s))-\nabla(V-V_k)(\gamma^j(s))}\:ds\\
		&\quad+\norm{\int_0^{\dt}\nabla V_k(q^j(s))\:ds}\,.
	\end{align*}
	By \eqref{q^j and Gamma^j}, the first term is $\O(\dt^3)$. For the second term we can use \eqref{q^j} and the observation that
		$$\int_0^{\dt}\nabla V_k(q^j(0)+sp_0)\:ds=0\,.$$
	Hence the second term is $\O(\dt^2)$. All in all, we obtain the estimate
	\begin{align*}
		\norm{p^j(\dt)-\dot{\gamma}^j(\dt)}=\O(\dt^2)\,.
	\end{align*}
	This estimate, together with \eqref{q^j and Gamma^j}, implies
	\begin{align}\label{vect Gamma}
		d\left(\vect{\gamma^{j+1}(0)}{\dot{\gamma}^{j+1}(0)},\vect{\gamma^j(\dt)}{\dot{\gamma}^j(\dt)}\right)=d\left(\vect{q^{j}(\dt)}{p^{j}(\dt)},\vect{\gamma^j(\dt)}{\dot{\gamma}^j(\dt)}\right)=\O(\dt^2)\,.
	\end{align}
	Since $\gamma^j$ and $\tilde{q}^j$ satisfy the same differential equation, say with associated Lipschitz constant $c$, Lemma \ref{lem:Gronwall} (with $f=g:(q,p)\mapsto(p,-\nabla(V-V_k)(q))$) implies that
	\begin{align}\label{Gamma and gamma time evolved}
		d\left(\vect{\gamma^j(t)}{\dot{\gamma}^j(t)},\vect{\tilde{q}^{j}(t)}{\tilde{p}^{j}(t)}\right)\leq e^{ct}d\left(\vect{\gamma^j(0)}{\dot{\gamma}^j(0)},\vect{\tilde{q}^j(0)}{\tilde{p}^j(0)}\right)\,.
	\end{align}
	Taking $t=\dt$, we by definition have
	\begin{align}\label{vect Gamma and gamma}
		d\left(\vect{\gamma^j(\dt)}{\dot{\gamma}^j(\dt)},\vect{\tilde{q}^{j+1}(0)}{\tilde{p}^{j+1}(0)}\right)\leq e^{c\dt}d\left(\vect{\gamma^j(0)}{\dot{\gamma}^j(0)},\vect{\tilde{q}^j(0)}{\tilde{p}^j(0)}\right)\,.
	\end{align}
	Combining \eqref{vect Gamma} and \eqref{vect Gamma and gamma}, we find that
	\begin{align*}
		d\left(\vect{\gamma^{j+1}(0)}{\dot{\gamma}^{j+1}(0)},\vect{\tilde{q}^{j+1}(0)}{\tilde{p}^{j+1}(0)}\right)\leq e^{c\dt}d\left(\vect{\gamma^j(0)}{\dot{\gamma^j}(0)},\vect{\tilde{q}^j(0)}{\tilde{p}^j(0)}\right)+\O(\dt^2)\,.
	\end{align*}
	Because $e^{jc\dt}=\O(1)$, repeated use of the above equation gives
	\begin{align}\label{Gamma^m and gamma^m}
		d\left(\vect{\gamma^m(0)}{\dot{\gamma}^m(0)},\vect{\tilde{q}^{m}(0)}{\tilde{p}^{m}(0)}\right)=\O(\dt)\,.
	\end{align}
	Let $t:=1-m\dt$. Using \eqref{Gamma and gamma time evolved}, we find
	\begin{align*}
		d\left(\vect{q(1)}{p(1)}, \vect{\tilde{q}(1)}{\tilde{p}(1)}\right)
		&\leq d\left(\vect{q(1)}{p(1)}, \vect{\gamma^m(t)}{\dot{\gamma}^m(t)}\right) + d\left(\vect{\gamma^m(t)}{
		\dot{\gamma}^m(t)}, \vect{\tilde{q}(1)}{\tilde{p}(1)}\right) \\
		&\leq d\left(q(1), \gamma^m(t)\right) + \|p(1) - \dot{\gamma}^m(t)\| \\
		&\quad+ e^{ct}
		d\left(\vect{\gamma^m(0)}{\dot{\gamma}^m(0)}, \vect{\tilde{q}^m(0)}{ \tilde{p}^m(0)}\right)\,.
	\end{align*}
	The first term is $\O(\dt^2)$ by \eqref{q^j and Gamma^j}, the second is $\O(\dt)$ by \eqref{p and Gamma dot}, and the last term is $\O(\dt)$ by \eqref{Gamma^m and gamma^m}. This implies \eqref{suffice to show big O}, and thereby the proposition.
\end{proof}
\noindent Proposition \ref{lem:long range convergence of dynamics} expresses a property of the classical time evolution associated to a trigonometric potential in terms of points in phase space. To translate this result to the world of observables, we fix $\epsilon>0$ and notice that any $g\in\Cr(T^*\T^n)$ is uniformly continuous. Hence for every $k\in\mN$ we may fix a $D_k$ such that 
	\begin{align} \label{tau_V*g-tau_(V-Vk)*g}
		\sup_{x\in U_k}|\Phi_V^*g(x)-\Phi_{V-V_k}^*g(x)|\leq \varepsilon\,,
	\end{align}
	where
		\[U_k:=\T^n\times\set{x\in\R^n}{ |k\cdot x|>D_k}\,.\]
	We also define the opens
	\begin{align*}
		W_k&:=\T^n\times\set{x\in\R^n}{ |k\cdot x|>2D_k}\,;\\
		U_\infty&:=\T^n\times\set{x\in\R^n}{ |k\cdot x|<4D_k\text{ for all }k\in\mN }\,;\\
		W_\infty&:=\T^n\times\set{x\in\R^n}{ |k\cdot x|<3D_k\text{ for all }k\in\mN }\,,
	\end{align*}
	and remark that $\{U_i\}_{i\in I}$ and $\{W_i\}_{i\in I}$ are open covers satisfying $\overline{W_i}\subseteq U_i$ for all $i\in I:=\mN\cup\{\infty\}$. 
	Since we already know how $\Phi_V^*g$ approximately behaves on $\bigcup_{k\in\mN}U_k$, let us see how it behaves on $U_\infty$.
	
	\begin{lem}\label{lem:f_infty}
		There exists an $f_\infty\in \Cr(T^*\T^n)$ that equals $\Phi_V^*g$ on $U_\infty$.
	\end{lem}	
	\begin{proof}
		Let $S:=\spn_\R~\mN$. We write our phase space as a product of topological spaces
			$$\T^n\times\R^n=(\T^n\times S)\times S^\perp\,,$$
		and note that
			$$C_0(\T^n\times S) \hatotimes \Wr(S^\perp)\,,$$
		is an ideal in $\Cr(T^*\T^n)$.
On the other hand, regarding our phase space as a coproduct of abelian Lie groups
			$$\T^n\times\R^n=(\T^n\times S) \oplus S^\perp\,,$$
		we define $\phi^t$ as the restriction of $\Phi^t_V$ to $\T^n\times S$ for each $t\in\R$. Because $\nabla V \perp S^\perp$, we have $\dot{p}(t)\perp S^\perp$, and hence
			$$\phi^t \colon\T^n\times S\rightarrow\T^n\times S\,$$
		is a well-defined homeomorphism. Moreover, we find the equation
		\begin{align*}
			\Phi^t_V(q,p_\|+p_\perp)&=\phi^t(q,p_\|)+(tp_\perp,p_\perp)\,,\quad\text{for all}\quad p_\|\in S,~p_\perp\in S^\perp\,,
		\end{align*}
		because its two sides solve the same differential equation.
%		The latter statement doesnt imply that the subalgebra
%			$$(\text{id}_{\T^n}\times P_S)^*(C_0(\T^n\times S))\subset \Cr(T^*\T^n)$$
%		is invariant under $\Phi_V$.
		Using the above relation in a straightforward calculation on generators, one can show that
			$$\Phi_V^*(C_0(\T^n\times S)\otimes\Wr(S^\perp))\subseteq C_0(\T^n\times S)\otimes\Wr(S^\perp)\,.$$
		Actually, the same holds for $\Phi_V^{-1}$, which implies that $\Phi_V^*$ is a *-automorphism of the ideal $C_0(\T^n\times S)\otimes\Wr(S^\perp)$. Now note that $U_\infty$ is of the form $K \times S^\perp$ for some compact subset $K \subseteq \T^n\times S$. By Urysohn's lemma, we may choose a function $\tilde{g}\in C_0(\T^n\times S)\otimes\Wr(S^\perp)$ that is $1$ on $U_\infty$, and define $f_\infty:=\tilde{g}\cdot\Phi_V^*g$. We then find that
			$$f_\infty=((\tilde{g}\circ\Phi_V^{-1})\cdot g)\circ\Phi_V\in C_0(\T^n\times S)\otimes\Wr(S^\perp)\,,$$
		and therefore $f_\infty\in\Cr(T^*\T^n)$.
	\end{proof}

\noindent We can finally prove that our classical resolvent algebra is invariant under any time evolution corresponding to a trigonometric potential.

\begin{prop}\label{prop: time evolution}
	For every trigonometric potential $V\colon\T^n\rightarrow\R$ and $g\in \Cr(T^*\T^n)$ we have $\Phi_V^*g\in\Cr(T^*\T^n)$.
\end{prop}
\begin{proof}
We use induction on the size of $\mN$ in $V=\sum_{k\in\mN} a_ke_k$ (while assuming that $\mN$ is chosen minimally). The induction base is precisely Lemma \ref{lem:free time evolution}.

We now carry out the induction step.
The induction hypothesis says that time evolution with respect to $V-V_k$ preserves $\Cr(T^*\T^n)$, for each $k\in\mN$. Therefore, writing $f_k:=\Phi_{V-V_k}^*g$, we have $f_k\in\Cr(T^*\T^n)$.
	Fixing $f_\infty$ as in Lemma \ref{lem:f_infty}, we have $f_i\in\Cr(T^*\T^n)$, and equation \eqref{tau_V*g-tau_(V-Vk)*g} implies that 
	\begin{align}\label{sup on U_i}
		\supnorm{f_i|_{U_i}-\Phi_V^*g|_{U_i}}<\epsilon\,,
	\end{align}
	for each $i\in I=\mN\cup\{\infty\}$.
	We now construct a partition of unity $\{\eta_i\}$ subordinate to the open cover $\{U_i\}$ of $\T^n \times \R^n$, to patch together the functions $\{f_i\}$ and obtain a single function in $\Cr(T^*\T^n)$.
	We start by defining nonnegative functions $\zeta_i\in \Cr(T^*\T^n)$ that are $1$ on $W_i$ and $0$ outside of $U_i$.
Explicitly, for each $k \in \mathcal{N}$, we take $\zeta_k := \mathbf{1}_{\T^n} \otimes (g_k\circ P_{\spn(k)})$ for some bump function $g_k$ on $\spn(k)$, and we take $\zeta_\infty := \mathbf{1}_{\T^n} \otimes (g_\infty \circ P_S)$ for some bump function $g_\infty$ on $S$. Because $\{W_i\}$ is a cover of $\T^n \times \R^n$, the sum $\sum_i\zeta_i\in\Cr(T^*\T^n)$ is bounded from below by 1, hence it is invertible in $\Cr(T^*\T^n)$, and therefore every function
		$$\eta_i:=\frac{\zeta_i}{\sum_j\zeta_j}\,,$$
	also lies in $\Cr(T^*\T^n)$. Now \eqref{sup on U_i} gives us $$\supnorm{\Phi_V^*g-\sum_i f_i\eta_i}<\varepsilon\,.$$
Since $\varepsilon > 0$ was arbitrary and $\Cr(T^*\T^n)$ is norm-closed, the assertion follows.
\end{proof}

\subsection{Arbitrary potentials}\label{sct:arbitrary potentials}

\noindent 
Having covered the trigonometric case, we now wish to tackle the general case.
The following lemma provides the required approximation of a generic potential by trigonometric ones.

\begin{lem}\label{lem:approximating nabla V}
	Let $V \in C^1(\T^n)$.
Then there exists a sequence $(V_m)_m$ of trigonometric polynomials such that $(\nabla V_m)_m$ converges uniformly to $\nabla V$.
Furthermore, if $V$ is real-valued, then every $V_m$ can be chosen to be real-valued as well.
\end{lem}

\begin{proof}
We construct the sequence $(V_m)$ by taking the convolution of $V$ with the $n$-dimensional analogues of the family of {\em Fej\'er kernels}.
We first recall that for each $m \geq 1$, the $m$-th Fej\'er kernel is given by
\begin{equation*}
F_{1,m} \colon \T \rightarrow \R\,, \quad 
q=[x] \mapsto \frac{1}{m} \sum_{k = 0}^{m - 1} \sum_{j = -k}^k e^{2\pi i j x}
= \frac{1}{m} \frac{\sin^2(\pi m x)}{\sin^2(\pi x)} \, ,
\end{equation*}
where the most right expression in this definition is understood to be equal to $m$ for $x = 0$.
The sequence $(F_{1,m})_{m \geq 1}$ is an approximation to the identity, i.e., for every continuous function $f$ on $\T$, the sequence $(F_{1,m} \ast f)_{m \geq 1}$ converges uniformly to $f$, where $\ast$ denotes the operation of convolution of functions \cite[sections 2.4 and 2.5.2]{stein03}.
%Moreover, $F_{1,m} \ast f$ is a trigonometric polynomial (on $\T$) for each $m \geq 1$.

Next, we define the $n$-dimensional analogues of these functions:
\begin{equation*}
F_{n,m} \colon \T^n \rightarrow \R\,, \quad 
q = (q_1,\ldots,q_n) \mapsto \prod_{l = 1}^n F_{1,m}(q_l) \, .
\end{equation*}
Using the corresponding fact for one-dimensional kernels, it is elementary to show that the sequence $(F_{n,m})_{m \geq 1}$ is an approximation to the identity.

We now define
\begin{equation*}
V_m := F_{n,m} \ast V \, ,
\end{equation*}
for each $m \geq 1$.
Because every $F_{n,m}$ is trigonometric, and $e_k*f=\hat{f}(k)e_k$ for every $f\in C(\T^n)$ and $k\in\Z^n$, the sequence $(V_m)_{m \geq 1}$ consists of trigonometric polynomials.
Moreover, by a general property of convolutions, we have
\begin{equation*}
\frac{\partial V_m}{\partial q_l}
= \frac{\partial}{\partial q_l} (F_{n,m} \ast V)
= F_{n,m} \ast \frac{\partial V}{\partial q_l}\,,
\end{equation*}
and since $(F_{n,m})_{m \geq 1}$ is an approximation to the identity, the right-hand side converges uniformly to $\frac{\partial V}{\partial q_l}$ as $m \to \infty$, for $l = 1,\ldots,n$.
It follows that $(\nabla V_m)_{m \geq 1}$ converges uniformly to $\nabla V$.
The final assertion is a consequence of the fact that the family of Fej\'er kernels (as well as its higher-dimensional analogues) consists of real-valued functions.
\end{proof}

\noindent We now extend Proposition \ref{prop: time evolution} to general $V$, thereby arriving at our final result.

\begin{thrm}
	Let $V\in C^1(\T^n)_\R$, and suppose that $\nabla V$ is Lipschitz continuous. Then we have 
		$$(\Phi^t_V)^*(\Cr(T^*\T^n))=\Cr(T^*\T^n)\,,$$
	for every $t\in\R$.
\end{thrm}

\begin{proof}
	It suffices to show that $(\Phi^t_V)^*(\Cr(T^*\T^n)) \subseteq \Cr(T^*\T^n)$; we can replace $t$ by $-t$ and note that $(\Phi_V^{-t})^\ast$ is the inverse of $(\Phi_V^t)^\ast$ to obtain the reverse inclusion.
By Lemma \ref{lem:t=1 or arbitrary t}, we may assume without loss of generality that $t = 1$.

Let $g \in \Cr(T^*\T^n)$.
By Lemma \ref{lem:approximating nabla V}, there exists a sequence of trigonometric potentials $(V_m)$ on $\T^n$ such that $(\nabla V_m)$ converges uniformly to $\nabla V$.
We show that this implies that $(\Phi_{V_m}^\ast (g))$ converges uniformly to $\Phi_V^\ast (g)$; since $\Phi_{V_m}^\ast (g) \in \Cr(T^*\T^n)$ by Proposition \ref{prop: time evolution} and since $\Cr(T^*\T^n)$ is norm-closed, the theorem will follow from this.

Let $\varepsilon > 0$, and let $c$ be the Lipschitz constant of $(q,p)\mapsto(p,-\nabla V(q))$.
Since $g$ is uniformly continuous, there exists $\delta > 0$ such that $|g(x) - g(y)| < \varepsilon$ for each $x,y \in \T^n \times \R^n$ with $d(x,y) < \delta$.
By assumption, there exists an $N \in \N$ such that for each $m \geq N$, we have $\|\nabla V - \nabla V_m\|_\infty < \delta e^{-c}$.
It follows from Lemma \ref{lem:Gronwall} that $d(\Phi_V(x),\Phi_{V_m}(x)) < \delta$ for each $x \in \T^n \times \R^n$ and each $m \geq N$, hence $\|\Phi_V^\ast(g) - \Phi_{V_m}^\ast(g)\|_\infty \leq \varepsilon$.
Thus $(\Phi_{V_m}^\ast (g))$ converges uniformly to $\Phi_V^\ast (g)$, as desired.
\end{proof}

\ifx\mycmd\undefined
	\phantomsection
	\addcontentsline{toc}{section}{References}
	\bibliographystyle{abbrv}
	\bibliography{./../Miscellaneous/References}

	\end{document}
\fi

\ifx\mycmd\undefined
	\documentclass[a4paper, 11pt, leqno]{article}
	
	\begin{document}
\fi

\section{Quantisation of the resolvent algebra}
\label{sec:quantisation}

%\subsection{Introduction}
%\label{subsec:quantum_resolvent_algebra_introduction}

\noindent 
Having shown the nice properties of $\Cr(T^*\T^n)$, we now ask whether there exists a quantum version of this algebra.
What complicates matters is that, contrary to the resolvent algebra $\mathcal{R}(\IR^{2n}, \sigma)$ of Buchholz and Grundling, on the cylinder it is hard, if not impossible, to define an algebra in terms of generators and relations implementing canonical commutation relations.
Thus we must take a different approach.

We will define our quantisation of the algebra $C_{\mathcal{R}}(T^\ast \IT^n)$ as an algebra represented on $L^2(\IT^n)$, using a version of Weyl quantisation similar to the definition of Landsman \cite[section II.3.4]{landsman98} for general Riemannian manifolds.
By contrast, Rieffel's algebras on cylinders in \cite{rieffel93}, apart from being quantisations of $C_u(T^\ast \IT^n)$ and subalgebras thereof, are in some sense universal objects from which a physical quantum system is obtained as the image of one of its irreducible representations, and it is not always clear which representation corresponds to the physical system that one wishes to model.
These algebras have many inequivalent irreducible representations due to the fact that $\IT$ is not simply connected, see e.g. \cite[Example 10.6]{rieffel93} and the discussion in \cite[section 7.7]{landsman17}.
By no means do we intend to discount such universal objects, however; we will return to this point in the outlook of this paper.
%in \cite{stienstra19}, one of the authors shows why such objects are likely better suited to fill the role of objects in the quantum category in the framework presented there.
The main advantage of quantising $C_{\mathcal{R}}(T^\ast \IT^n)$ as an algebra of operators on $L^2(\IT^n)$ lies in the explicit formula for the quantisations of the generators of $C_{\mathcal{R}}(T^\ast \IT^n)$ that we are able to derive.

This section is structured as follows.
In subsection \ref{subsec:definition_of_quantisation_map}, we define the Weyl quantisation map and prove the aforementioned explicit formula.
In subsection \ref{subsec:quantisation_is_strict}, we show that, except for continuity of the map $\hbar \mapsto \|\mathcal{Q}^W_\hbar(f)\|$ at $\hbar > 0$ for fixed $f \in C_{\mathcal{R}}(T^\ast \IT^n)$, the quantisation is strict.

\subsection{Definition of the quantisation map}
\label{subsec:definition_of_quantisation_map}

\noindent 
Let us first recall the basics of Weyl quantisation in $\IR^{2n}$, the quantisation procedure in \cite{weyl27} conceived by Weyl.
Given say, a Schwartz function $f \in \mathcal{S}(\IR^{2n})$, one associates an operator $\mathcal{Q}^W_\hbar(f) \in B(L^2(\IR^n))$ to it as follows.
First, one expresses $f$ in terms of functions of the form
\begin{equation*}
\IR^{2n} = \IR^n \times \IR^n \rightarrow \IC, \quad 
(q,p) \mapsto e^{i(a \cdot q + b \cdot p)}\,,
\end{equation*}
where $a,b \in \IR^n$, by considering the Fourier transform of $f$.
One subsequently substitutes these exponential functions with the operators
\begin{equation*}
e^{i(a \cdot Q + b \cdot P)}\,,
\end{equation*}
where $Q,P$ are vectors whose components are the essentially self-adjoint operators on $\S(\R^n)\subseteq L^2(\R^n)$, defined by $Q_j\psi(x):=x_j\psi(x)$ and $P_j\psi(x):=-i\hbar\frac{d\psi}{dx_j}(x)$.
Thus, the Weyl quantisation of a function $f$ is informally given by the expression
\begin{align*}
&(2\pi)^{-2n} \int_{\IR^n} \int_{\IR^n} \int_{\IR^n} \int_{\IR^n} f(q,p) e^{ia \cdot (Q - q) + ib \cdot (P - p)} \: dq \: dp \: da \: db \\
&\quad = (2\pi)^{-2n} \int_{\IR^n} \int_{\IR^n} \int_{\IR^n} \int_{\IR^n} f(q,p) e^{i\hbar\frac{a \cdot b}{2}} e^{ia \cdot (Q - q)} e^{ib \cdot (P - p)} \: dq \: dp \: da \: db\,,
\end{align*}
where we take $\hbar > 0$.
To define the above integrals rigorously, we can insert a function $\psi \in \mathcal{S}(\IR^n)$ on the right-hand side of the integrand, and check that the resulting expression is well-defined and that it defines a bounded operator on $\mathcal{S}(\IR^n)$ viewed as a subspace of $L^2(\IR^n)$.
Since $\mathcal{S}(\IR^n)$ is dense in $L^2(\IR^n)$, the operator has a unique bounded extension to $L^2(\IR^n)$, which we define to be $\mathcal{Q}^W_\hbar(f)$.
Using standard identities for Fourier transforms of functions, and performing a number of substitutions, it can be shown that
\begin{equation*}
(\mathcal{Q}^W_\hbar(f)\psi)(x)
= (2 \pi \hbar)^{-n} \int_{\IR^n} \int_{\IR^n} f\left(x + \frac{y}{2}, p \right) e^{-i\frac{ y \cdot p}{\hbar}} \psi(x + y) \: dp \: dy\,,
\end{equation*}
for each $\psi \in \mathcal{S}(\IR^n)$ and each $x \in \IR^n$.

We now adapt the Weyl quantisation formula to $T^\ast \IT^n$ in such a way that we can quantise elements of $C_{\mathcal{R}}(T^\ast \IT^n)$.
We already identified a dense Poisson algebra of $\Cr(T^*\T^n)$ in Section \ref{sct:Basic Results}, namely the space $\mathcal{S}_{\mathcal{R}}(T^\ast \IT^n)$ of finite linear combinations of functions of the form $e_k \otimes h_{U,\xi,g}$; see Proposition \ref{prop:Poisson_subalg_for_arbitrary_n}.
These are the functions that we will quantise.
To handle such functions, we take inspiration from Rieffel's work \cite{rieffel93}, regarding the integrals in the above formula as oscillatory integrals, and regularising the expression by inserting a factor in the integrand in the form of a member of a net of functions that converges pointwise to the constant function $1_{\R^n}$, as in part (1) of the next proposition.
Part (2) of this proposition is the analogue of \cite[Proposition 1.11]{rieffel93}.

\begin{prop}

\noindent 
\begin{myenum}
\item Let $f \in \mathcal{S}_{\mathcal{R}}(T^\ast \IT^n)$, let $\hbar > 0$, and let $\psi \in C(\IT^n)$.
Then for each $[x] \in \IT^n$, the limit
\begin{equation}
\lim_{\delta \to 0} (2 \pi \hbar)^{-n} \int_{\IR^n} \int_{\IR^n} f\left(\left[x + \tfrac{1}{2}y\right], p \right) e^{-\frac{\delta}{2}p^2} e^{-i\frac{ y \cdot p}{\hbar}} \psi[x + y] \: dp \: dy\,,
\label{eq:Weyl_quantisation_formula}
\end{equation}
exists.
\item Now assume $f = e_k \otimes h_{U,\xi,g}$ is a function as described in Definition \ref{def:resolvent_algebra_Schwartz_functions}.
Then the expression in equation \eqref{eq:Weyl_quantisation_formula} is equal to
\begin{equation*}
(2 \pi \hbar)^{-\dim(U)} e^{\pi i k \cdot \hbar\xi}e^{2 \pi i k \cdot x}
\int_U \int_U g\left(p + \pi \hbar P_U(k) \right) e^{-i\frac{ y \cdot p}{\hbar}} \psi[x + y + \hbar \xi] \: dp \: dy\,.
\end{equation*}
\end{myenum}
For each $l \in \IZ^n$, let $\psi_l$ be the function
\begin{equation*}
\IT^n \rightarrow \IC, \quad 
[x] \mapsto e^{2 \pi i l \cdot x}\,,
\end{equation*}
and regard it as an element of $L^2(\IT^n)$.
\begin{myenum}
\setcounter{enumi}{2}
\item In addition to the assumptions in the previous part of the proposition, suppose that $\psi = \psi_l$ for some $l \in \IZ^n$.
Then the expression in equation \eqref{eq:Weyl_quantisation_formula} is equal to
\begin{equation*}
h_{U,\xi,g}(\pi \hbar(k + 2l)) \psi_{k + l}[x]\,,
\end{equation*}
and the map defined on $\spn_{l\in\Z^n}\{\psi_l\}$ sending $\psi$ to the function on $\IT^n$ that assigns to a point $[x] \in \IT^n$ the limit in \eqref{eq:Weyl_quantisation_formula} extends in a unique way to a bounded linear operator on $L^2(\IT^n)$ with norm $\leq \|g\|_\infty$.
\end{myenum}
\label{prop:Weyl_quantisation_is_well-defined}
\end{prop}

\begin{proof}
We first show that for functions $f$ of the form $e_k \otimes h_{U,\xi,g}$, i.e., $f$ as in part (2) of the proposition, the limit in equation \eqref{eq:Weyl_quantisation_formula} exists, and is equal to the formula in part (2) of the proposition.
Since $\mathcal{S}_{\mathcal{R}}(T^\ast \IT^n)$ is by definition the linear span of such functions, part (1) will follow from this. Thus, take such an $f$, and note that for any $\delta > 0$, we have
\begin{align*}
&(2 \pi \hbar)^{-n} \int_{\IR^n} \int_{\IR^n} f\left([x + \tfrac{1}{2}y], p \right) e^{-\frac{\delta}{2}p^2} e^{-i\frac{ y \cdot p}{\hbar}} \psi[x + y ] \: dp \: dy \\
&\quad= (2 \pi \hbar)^{-n} \int_{\IR^n} \int_{\IR^n}  e^{i \left(\xi \cdot p - \frac{y \cdot p}{\hbar}\right)} g \circ P_U(p) e^{-\frac{\delta}{2}p^2} \: dp \: e^{2 \pi i k \cdot \left(x + \frac{y}{2}\right)} \psi[x + y ]  \: dy \\
&\quad= (2 \pi \hbar)^{-n} \int_{\IR^n} \int_{\IR^n} e^{-i \frac{y \cdot p}{\hbar}} g \circ P_U(p) e^{-\frac{\delta}{2}p^2} \: dp \: e^{2 \pi i k \cdot \left(x + \frac{y + \hbar \xi}{2}\right)} \psi[x + y + \hbar \xi ]  \: dy\,.
\end{align*}
The inner integral over $p$ can be written as a product of two integrals; one over $U$ and one over $U^\perp$:
\begin{align*}
&\int_{\IR^n} e^{-i \frac{y \cdot p}{\hbar}} g \circ P_U(p) e^{-\frac{\delta}{2}p^2} \: dp \\
&\quad = \int_U g(p_1) e^{-\frac{\delta}{2}p_1^2} e^{-i \frac{P_U(y) \cdot p_1}{\hbar}} \: dp_1 \cdot \int_{U^\perp} e^{-\frac{\delta}{2}p_2^2} e^{-ip_2 \cdot \frac{y - P_U(y)}{\hbar}} \: dp_2 \\
&\quad = \int_U g(p_1) e^{-\frac{\delta}{2}p_1^2} e^{-i \frac{P_U(y) \cdot p_1}{\hbar}} \: dp_1 \cdot (2 \pi \delta^{-1})^{\frac{\dim(U^\perp)}{2}} e^{-\frac{1}{2\delta \hbar^2}(y - P_U(y))^2}\,.
\end{align*}
Inserting this back into the previous displayed formula, and splitting the outer integral in that formula into an integral over $U$ and an integral over $U^\perp$, we obtain
\begin{align*}
&(2 \pi \hbar)^{-n} \int_{\IR^n} \int_{\IR^n} f\left([x + \tfrac{1}{2}y], p \right) e^{-\frac{\delta}{2}p^2} e^{-i\frac{ y \cdot p}{\hbar}} \psi[x + y ] \: dp \: dy \\ 
&\quad = (2 \pi \hbar)^{-\dim(U)} \int_{U} h_{1,\delta}(y_1) \int_{U^\perp} h_{2,\delta}(y_1,y_2) \: dy_2 \: dy_1\,,
\end{align*}
where
\begin{align*}
h_{1,\delta} \colon U &\rightarrow \IC, \\
y_1 &\mapsto e^{2 \pi i k \cdot \left(x + \frac{y_1 + \hbar \xi}{2}\right)}
\int_U g(p_1) e^{-\frac{\delta}{2}p_1^2} e^{-i \frac{y_1 \cdot p_1}{\hbar}} \: dp_1\,,
\end{align*}
and 
\begin{align*}
h_{2,\delta} \colon U \times U^\perp &\rightarrow \IC, \\
(y_1, y_2) &\mapsto (2 \pi \delta \hbar^2)^{\frac{-\dim(U^\perp)}{2}} e^{-\frac{1}{2\delta \hbar^2}y_2^2} \cdot \psi[x + y_1 + y_2 + \hbar \xi ] e^{\pi i k \cdot y_2}\,.
\end{align*}
Now note that the family of functions
\begin{equation*}
U^\perp \rightarrow \IR, \quad 
y_2 \mapsto (2 \pi \delta \hbar^2)^{\frac{-\dim(U^\perp)}{2}} e^{-\frac{1}{2\delta \hbar^2}y_2^2}\,,
\end{equation*}
indexed by $\delta > 0$ is an approximation to the identity for functions on $U^\perp$.
By continuity of $\psi$, it follows that the functions
\begin{equation*}
h_{3,\delta} \colon U \rightarrow \IC, \quad 
y_1 \mapsto \int_{U^\perp} h_{2,\delta}(y_1,y_2) \: dy_2 \,,
\end{equation*}
converge pointwise to the function
\begin{equation*}
U \rightarrow \IC, \quad 
y_1 \mapsto \psi[x + y_1 + \hbar \xi ]\,,
\end{equation*}
as $\delta \to 0$.
Moreover, they are bounded, with $\|h_{3,\delta}\|_\infty \leq \|\psi\|_\infty$ for each $\delta > 0$.
In addition, by the dominated convergence theorem, the functions $h_{1,\delta}$ converge pointwise to the function
\begin{equation*}
U \rightarrow \IC, \quad 
y_1 \mapsto e^{2 \pi i k \cdot \left(x + \frac{y_1 + \hbar \xi}{2}\right)} \int_U g(p_1) e^{-i \frac{y_1 \cdot p_1}{\hbar}} \: dp_1\,,
\end{equation*}
as $\delta \to 0$.
Indeed, the integrands defining these functions are all dominated by the integrable function $|g|$.
Furthermore, note that
\begin{align*}
&\int_U g(p_1) e^{-\frac{\delta}{2} p_1^2} e^{-i \frac{y_1 \cdot p_1}{\hbar}} \: dp_1 \\
&\quad= \frac{(1 + \|y_1\|^2)^{\dim(U)}}{(1 + \|y_1\|^2)^{\dim(U)}} \int_U g(p_1) e^{-\frac{\delta}{2} p_1^2} e^{-i \frac{y_1 \cdot p_1}{\hbar}} \: dp_1 \\
&\quad= \frac{1}{(1 + \|y_1\|^2)^{\dim(U)}} \int_U (1 - \hbar^2 \Delta_U)^{\dim(U)}(g(p^\prime) e^{-\frac{\delta}{2} (p^\prime)^2})|_{p^\prime = p_1} e^{-i \frac{y_1 \cdot p_1}{\hbar}} \: dp_1\,,
\end{align*}
where $\Delta_U$ denotes the standard Laplacian on $U$, and that for the family of the functions
\begin{equation*}
U \rightarrow \IC, \quad 
p_1 \mapsto (1 - \hbar^2 \Delta_U)^{\dim(U)}(g(p^\prime) e^{-\frac{\delta}{2} (p^\prime)^2})|_{p^\prime = p_1}\,,
\end{equation*}
indexed by $\delta \in (0,C]$, where $C$ is an arbitrary positive real number, there exists a positive function $H_C \in L^1(U)$ dominating the entire family.
It follows that for each $\delta \in (0,C]$ and each $y_1 \in U$, we have
\begin{equation*}
|h_{1,\delta}(y_1)| \leq \frac{\|H_C\|_1}{(1 + \|y_1\|^2)^{\dim(U)}}\,.
\end{equation*}
The (absolute values of the) functions
\begin{equation*}
U \rightarrow \IC, \quad 
y_1 \mapsto h_{1,\delta}(y_1) \int_{U^\perp} h_{2,\delta}(y_1,y_2) \: dy_2\,,
\end{equation*}
are therefore dominated by the integrable function
\begin{equation*}
y_1 \mapsto \frac{\|H_C\|_1 \|\psi\|_\infty}{(1 + \|y_1\|^2)^{\dim(U)}}\,,
\end{equation*}
so we may again invoke the dominated convergence theorem to find that
\begin{align*}
&\lim_{\delta \to 0} (2 \pi \hbar)^{-\dim(U)} \int_{U} h_{1,\delta}(y_1) \int_{U^\perp} h_{2,\delta}(y_1,y_2) \: dy_2 \: dy_1 \\
&\quad= (2 \pi \hbar)^{-\dim(U)} \int_{U} \left( \lim_{\delta \to 0} h_{1,\delta}(y_1) \right) \left( \lim_{\delta \to 0}\int_{U^\perp} h_{2,\delta}(y_1,y_2) \: dy_2 \right) \: dy_1 \\
&\quad= (2 \pi \hbar)^{-\dim(U)} \int_U \int_U g(p_1) e^{-i \frac{y_1 \cdot p_1}{\hbar}} \: dp_1 \:
e^{2 \pi i k \cdot \left(x + \frac{y_1 + \hbar \xi}{2}\right)} \psi[x + y_1 + \hbar \xi ] \: dy_1 \\
&\quad= \frac{e^{\pi i k \cdot \hbar \xi} e^{2 \pi i k \cdot x}}{(2 \pi \hbar)^{\dim(U)}} \int_U \int_U g(p_1) e^{-i y_1 \cdot \left(\frac{p_1}{\hbar} - \pi k \right)} \: dp_1 \: \psi[x + y_1 + \hbar \xi ] \: dy_1 \\
&\quad= \frac{e^{\pi i k \cdot \hbar \xi} e^{2 \pi i k \cdot x}}{(2 \pi \hbar)^{\dim(U)}} \int_U \int_U g(p_1 + \pi \hbar P_U(k)) e^{-i \frac{y_1 \cdot p_1}{\hbar}} \: dp_1 \: \psi[x + y_1 + \hbar \xi ] \: dy_1\,,
\end{align*}
which completes our proof of part (2).\\

\noindent 
For part (3), we simply take $\psi = \psi_l \in C(\IT^n) \subset L^2(\IT^n)$, with $l \in \IZ^n$, and apply the formula we just found:
\begin{align*}
&(2 \pi \hbar)^{-\dim(U)} e^{\pi i k \cdot \hbar \xi} e^{2 \pi i k \cdot x} \int_U \int_U g(p_1 + \pi \hbar P_U(k)) e^{-i \frac{y_1 \cdot p_1}{\hbar}} \: dp_1 \: e^{2 \pi i l \cdot (x + y_1 + \hbar \xi)} \: dy_1 \\
&\quad= (2 \pi \hbar)^{-\dim(U)} e^{\pi i (k + 2l) \cdot \hbar \xi} e^{2 \pi i (k + l) \cdot x} \int_U \int_U g(p_1 + \pi \hbar P_U(k)) e^{-i y_1 \cdot \left( \frac{p_1}{\hbar} - 2\pi l\right)} \: dp_1 \: dy_1 \\
&\quad= (2 \pi)^{-\dim(U)} e^{\pi i (k + 2l) \cdot \hbar \xi} e^{2 \pi i (k + l) \cdot x} \int_U \int_U g(p_1 + \pi \hbar P_U(k + 2l)) e^{-i y_1 \cdot p_1} \: dp_1 \: dy_1 \\
&\quad= e^{\pi i (k + 2l) \cdot \hbar \xi} e^{2 \pi i (k + l) \cdot x} g \circ P_U(\pi \hbar(k + 2l)) \\
&\quad= h_{U,\xi,g}(\pi \hbar(k + 2l)) \psi_{k + l}[x]\,,
\end{align*}
which proves the formula in part (3).

We thus see that the linear map on $\spn_l\{\psi_l\}$ uniquely determined by
\begin{equation*}
\psi_l  \mapsto h_{U,\xi,g}(\pi \hbar(k + 2l)) \psi_{k + l}\,,
\end{equation*}
maps an orthonormal basis to an orthogonal system of vectors in $L^2(\IT^n)$, and the norm of the image of such a vector $\psi_l$ is less than or equal to $\|g\|_\infty = \|h_{U,\xi,g}\|_\infty = \|f\|_\infty$.
(Note that the suprema defining these sup-norms are taken over $U$, $\IR^n$ and $\IT^n \times \IR^n$, respectively.)
Because of this and the fact that the $\psi_l$'s densely span $L^2(\IT^n)$, the map extends in a unique way to a bounded operator on $L^2(\IT^n)$ with norm $\leq \|g\|_\infty$, which proves the final assertion.
\end{proof}

\noindent 
The proposition justifies the following definitions:
\begin{defi}
For each $\hbar > 0$ and each $f \in \mathcal{S}_{\mathcal{R}}(T^\ast \IT^n)$, we define the {\em Weyl quantisation $\mathcal{Q}^W_\hbar(f)$ of $f$} to be the unique bounded linear extension of the operator on $\spn_{l\in\Z^n}\{\psi_l\}$ defined by the formula
\begin{align*}
(\mathcal{Q}^W_\hbar(f) \psi)[x] 
:= \lim_{\delta \to 0} (2 \pi \hbar)^{-n} \int_{\IR^n} \int_{\IR^n} f\left([x + \tfrac{1}{2}y], p \right) e^{-\frac{\delta}{2}p^2} e^{-i\frac{ y \cdot p}{\hbar}} \psi[x + y] \: dp \: dy\,.
\end{align*}
We thus obtain a map, the {\em Weyl quantisation map $\mathcal{Q}^W_\hbar \colon \mathcal{S}_{\mathcal{R}}(T^\ast \IT^n) \rightarrow B(L^2(\IT^n))$}, for each $\hbar > 0$.
We define the {\em quantum resolvent algebra $A_\hbar$ on $\IT^n \times \IR^n$} to be the C$^\ast$-subalgebra of $B(L^2(\IT^n))$ generated by the image of $\Sr(\T^*\T^n)$ under $\mathcal{Q}^W_\hbar$.
\end{defi}

\noindent
Part (3) of Proposition \ref{prop:Weyl_quantisation_is_well-defined} can now be phrased as an explicit formula for the Weyl quantisation of a generator $e_k\otimes h\in\Sr(T^*\T^n)$, namely
	\begin{equation}\label{eq:formula}
		\QW(e_k\otimes h)\psi_l=h(\pi\hbar(k+2l))\psi_{k+l}\,.
	\end{equation}

\begin{prop}
Let $\hbar > 0$.
\begin{myenum}
\item The Weyl quantisation map is linear and *-preserving;

\item For each $\hbar^\prime > 0$, we have $A_\hbar = A_{\hbar^\prime}$;

\item The image of
\begin{equation*}
\text{\normalfont span}_\IC \set{e_k \otimes g}{k \in \IZ^n, \: g \in \mathcal{S}(\IR^n)}
\subseteq \mathcal{S}_{\mathcal{R}}(T^\ast \IT^n) \cap C_0(T^\ast \IT^n) \,,
\end{equation*}
under $\mathcal{Q}^W_\hbar$ is a dense subspace of $B_0(L^2(\IT^n))$;

\item Under the canonical embedding
\begin{equation*}
B(L^2(\T^n))\hookrightarrow B(L^2(\T^{n+m})) \cong B(L^2(\T^n)) \hatotimes B(L^2(\T^m)) \,, \quad a \mapsto a \otimes \mathbf{1} \,,
\end{equation*}
induced by the projection at the level of configuration spaces $\T^{n+m}\rightarrow\T^{n}$ onto the first $n$ coordinates, the image of the quantum resolvent algebra on $T^\ast \IT^{n + m}$ is a subalgebra of the quantum resolvent algebra on $T^\ast \IT^{n}$.
(Here, $\hatotimes$ denotes the von Neumann algebraic tensor product.)

\item Let $\rho_0$ be the group representation of $\IT^n$ on $C_b(T^\ast \IT^n)$ given by
\begin{align*}
%(\rho_0[x]f)(q,p) := f(-x+q,p)
\rho_0[x]f:=\left(\,(q,p) \mapsto f(-x+q,p)\,\right)\,,
\end{align*}
and let $\rho_\hbar$ be the group representation of $\IT^n$ on $B(L^2(\IT^n))$ given by
\begin{equation*}
\rho_\hbar[x]a
:=  L[x] a L[-x]\,,
\end{equation*}
where $L \colon \IT^n \rightarrow U(L^2(\IT^n))$ denotes the left regular representation of $\IT^n$.
Then both $C_{\mathcal{R}}(T^\ast \IT^n)$ and $\mathcal{S}_{\mathcal{R}}(T^\ast \IT^n)$ are invariant under $\rho_0$.
Furthermore, the Weyl quantisation map is equivariant with respect to these representations.
\end{myenum}
\label{prop:quantisation_notable_properties}
\end{prop}

\begin{rem}
Because of part (2) of this proposition, we will write $A_\hbar$ for the C$^\ast$-algebra generated by $\mathcal{Q}^W_{\hbar^\prime}(\mathcal{S}_{\mathcal{R}}(T^\ast \IT^n))$ for any value of $\hbar^\prime > 0$ without specifying $\hbar$.
Part (3) is the analogue of the first part of \cite[Corollary II.2.5.4]{landsman98} in the present setting, while part (5) is the analogue of \cite[Theorem II.2.5.1]{landsman98}.
\end{rem}

\begin{proofenum}
\item Linearity of $\mathcal{Q}^W_\hbar$ is obvious from the definition.
Now let $e_k \otimes h$ be a generator of $\mathcal{S}_{\mathcal{R}}(T^\ast \IT^n)$, and let
\begin{equation*}
\mathcal{F} \colon L^2(\IT^n) \rightarrow \ell^2(\IZ^n), \quad 
\psi^\prime \mapsto (\,a \mapsto \inp{\psi_a}{\psi^\prime}\,)\,,
\end{equation*}
be the Fourier transform.
Here, $\inp{\cdot}{\cdot}$ denotes the usual inner product on $L^2(\IT^n)$.
We follow the physicists' convention, taking the inner product to be linear in its second argument.
It follows from \eqref{eq:formula} that 
\begin{equation*}
\mathcal{Q}^W_\hbar(e_k \otimes h)
= \mathcal{F}^{-1} S^k M_{h_1} \mathcal{F}\,,
\end{equation*}
where $S^k\colon\ell^2(\IZ^n) \rightarrow \ell^2(\IZ^n)$ denotes the shift operator defined by
\begin{equation*}
(S^k\phi)(l):= \phi(l-k)\,,
\end{equation*}
and $M_{h_1}$ denotes the multiplication operator on $\ell^2(\IZ^n)$ associated to the function
\begin{equation*}
h_1 \colon \IZ^n \rightarrow \IC, \quad 
l \mapsto h(\pi \hbar(k + 2l))\,.
\end{equation*}
Next, for each $l \in \IZ^n$, we have
\begin{align*}
(S^k M_{h_1})^\ast \delta_l
&= M_{\overline{h_1}} S^{-k} \delta_l
= \overline{h}(\pi \hbar(k + 2(l - k))) \delta_{l - k} \\
&= \overline{h}(\pi \hbar(-k + 2l)) \delta_{l - k}
= S^{-k} M_{h_2} \delta_l\,,
\end{align*}
where $h_2$ is defined as $h_2(l):=\overline{h}(\pi \hbar(-k + 2l))$.
Also note that
\begin{equation*}
\mathcal{Q}^W_\hbar(\overline{e_k \otimes h})
= \mathcal{Q}^W_\hbar(e_{-k} \otimes \overline{h})
= \mathcal{F}^{-1} S^{-k} M_{h_2} \mathcal{F}\,,
\end{equation*}
so by unitarity of the Fourier transform, we have
\begin{equation*}
\mathcal{Q}^W_\hbar(\overline{e_k \otimes h})
= \mathcal{F}^{-1} (S^k M_{h_1})^\ast \mathcal{F}
= (\mathcal{F}^{-1} S^k M_{h_1} \mathcal{F})^\ast
= \mathcal{Q}^W_\hbar(e_k \otimes h)^*\,,
\end{equation*}
hence $\mathcal{Q}^W_\hbar$ is indeed compatible with the involutions.

\item For each $\hbar > 0$, each $f \in \mathcal{S}_{\mathcal{R}}(T^\ast \IT^n)$ and each $\psi \in L$, we have
\begin{align*}
(\mathcal{Q}^W_\hbar(f) \psi)(x)
 = \lim_{\delta \to 0} (2 \pi)^{-n} \int_{\IR^n} \int_{\IR^n} f\left([x + \tfrac{1}{2}y], \hbar p^\prime \right) e^{-\frac{\delta}{2}(p^\prime)^2} e^{-iy \cdot p^\prime} \psi[x + y] \: dp^\prime \: dy\,,
\end{align*}
where we have made the substitution $p = \hbar p^\prime$ in the formula defining $\mathcal{Q}^W_\hbar(f) \psi$, and absorbed a factor $\hbar^2$ in $\delta$.
Next, we observe that $\mathcal{S}_{\mathcal{R}}(T^\ast \IT^n)$ is closed under the map
\begin{equation*}
f \mapsto (\,(q,p) \mapsto f(q, Cp)\,),
\end{equation*}
for each $C \in \IR$, in particular for $C = \hbar^\prime / \hbar$ for any $\hbar, \hbar^\prime > 0$.
It follows that $\mathcal{Q}^W_\hbar(\mathcal{S}_{\mathcal{R}}(T^\ast \IT^n)) = \mathcal{Q}^W_{\hbar^\prime}(\mathcal{S}_{\mathcal{R}}(T^\ast \IT^n))$, hence $A_\hbar = A_{\hbar^\prime}$, as desired.

\item Let $B$ be the left-hand side of the displayed formula in the statement.
Now let $k \in \IZ^n$, and let $g \in \mathcal{S}(\IR^n)$.
Using notation from the proof of part (1) of this proposition, we have
\begin{equation*}
\mathcal{Q}^W_{\hbar}(e_k \otimes g)
= \mathcal{F}^{-1} S^k M_{g_1} \mathcal{F}\,,
\end{equation*}
where $g_1$ denotes the function
\begin{equation*}
\IZ^n \rightarrow \IC, \quad 
l \mapsto g(\pi\hbar(k + 2l))\,.
\end{equation*}
This function vanishes at infinity, so its corresponding multiplication operator $M_{g_1}$ is compact.
All of the other operators that we compose to obtain $\mathcal{Q}^W_{\hbar}(e_k \otimes g)$ are bounded, hence $\mathcal{Q}^W_{\hbar}(e_k \otimes g)$ is compact.
Since $\mathcal{Q}^W_\hbar$ is a linear map and $B_0(L^2(\IT^n))$ is a linear subspace of $B(L^2(\IT^n))$, it follows that $\mathcal{Q}^W_\hbar(B) \subseteq B_0(L^2(\IT^n))$.

To prove the assertion that $\mathcal{Q}^W_\hbar(B)$ is in fact a dense subspace of $B_0(L^2(\IT^n))$, we note that, given $a$ and $b$ in $\IZ^n$, we can fix a $g \in \mathcal{S}(\IR^n)$ such that
\begin{equation*}
g(\pi\hbar(a - b + 2l)) = \delta_{l,b}\,,
\end{equation*}
for each $l \in \IZ^n$.
It follows that, in bra-ket notation,
\begin{equation*}
\mathcal{Q}^W_\hbar(e_{a - b} \otimes g) = |\psi_a \rangle \langle \psi_b|\,,
\end{equation*}
and from the fact that $a,b \in \IZ^n$ were arbitrary and that the family of vectors $(\psi_l)_{l \in \IZ^n}$ is an orthonormal basis of $L^2(\IT^n)$, we infer that $\mathcal{Q}^W_\hbar(B)$ is dense in $B_0(L^2(\IT^n))$.

\item From Definition \ref{def:resolvent_algebra_Schwartz_functions} one straightforwardly shows that $\Sr(T^*\T^n)\otimes\C\unit_{\T^m\times\R^m}\subseteq\Sr(T^*\T^{n+m})$. From formula \eqref{eq:formula}, one obtains $\QW(f\otimes\unit_{\T^m\times\R^m})=\QW(f)\otimes\mathbbm{1}$ for all $f\in\Sr(T^*\T^n)$. Therefore, $$\QW(\Sr(T^*\T^n))\otimes\mathbbm{1}\subseteq\QW(\Sr(T^*\T^{n+m}))\,,$$ which implies the same inclusion for the respective generated C*-algebras.

\item Suppose $f$ is of the form $e_k \otimes h$.
Then it is readily seen that
\begin{equation*}
\rho_0[x](e_k \otimes h)
= e^{-2 \pi i k \cdot x} e_k \otimes h \in \mathcal{S}_{\mathcal{R}}(T^\ast \IT^n)\,,
\end{equation*}
for each $[x] \in \IT^n$, from which it follows that both $C_{\mathcal{R}}(T^\ast \IT^n)$ and $\mathcal{S}_{\mathcal{R}}(T^\ast \IT^n)$ are invariant subspaces of the representation $\rho_0$.
Furthermore, for each $l \in \IZ^n$, we have
\begin{align*}
(\rho_\hbar[x](\mathcal{Q}^W_\hbar(e_k \otimes h)))\psi_l 
& = L[x] \mathcal{Q}^W_\hbar(e_k \otimes h) L[-x] \psi_l \\
& = e^{2 \pi i l \cdot x} L[x] \mathcal{Q}^W_\hbar(e_k \otimes h) \psi_l \\
& = e^{2 \pi i l \cdot x} h(\pi \hbar(k + 2l)) L[x] \psi_{k + l} \\
& = e^{2 \pi i l \cdot x} e^{-2 \pi i (k + l) \cdot x} h(\pi \hbar(k + 2l)) \psi_{k + l} \\
& = \mathcal{Q}^W_\hbar( e^{-2 \pi i k \cdot x} e_k \otimes h) \psi_l\,,
\end{align*}
from which we conclude that
\begin{equation*}
\rho_\hbar[x](\mathcal{Q}^W_\hbar(e_k \otimes h))
= \mathcal{Q}^W_\hbar( \rho_0[x](e_k \otimes h))\,,
\end{equation*}
for each $[x]$ and each generator $e_k \otimes h$ of $\mathcal{S}_{\mathcal{R}}(T^\ast \IT^n)$.
Since these generators span $\mathcal{S}_{\mathcal{R}}(T^\ast \IT^n)$, and the quantisation map and the maps $\rho_0[x]$ and $\rho_\hbar[x]$ are linear, we may substitute for $e_k \otimes h$ any element of $\mathcal{S}_{\mathcal{R}}(T^\ast \IT^n)$ in the above equation.
\end{proofenum}

\subsection{Proof of strict quantisation}
\label{subsec:quantisation_is_strict}

\noindent 
We now show that Weyl quantisation as defined in the previous section yields a strict quantisation of the dense Poisson subalgebra $\mathcal{S}_{\mathcal{R}}(T^\ast \IT^n)$ of the classical resolvent algebra $C_{\mathcal{R}}(T^\ast \IT^n)$ on $T^\ast \IT^n \cong \IT^n \times \IR^n$, see \cite[section II.1.1.1]{landsman98} or Theorem \ref{thrm:Weyl_quantisation_is_strict} below.
Of these properties, the most difficult one to prove is Rieffel's condition, i.e., convergence of the operator norms of $\mathcal{Q}^W_\hbar(f)$ to the sup-norm of $f \in \mathcal{S}_{\mathcal{R}}(T^\ast \IT^n)$, which we discuss separately before showing that the other conditions hold.
To prepare for the proof, we first make the following observation:

\begin{lem}
For $K\in\N\backslash\{0\}$ let $K\IZ^n := K \IZ \times \dots \times K \IZ$, and let $\IZ^n_K := \IZ^n / K\IZ^n$.
For each $k \in \IZ^n_K$, let $S_{\text{\normalfont per}}^k \colon \ell^2(\IZ^n_K) \rightarrow \ell^2(\IZ^n_K)$ be the operator given by
\begin{equation*}
\phi \mapsto (\,l \mapsto \phi(-k + l)\,) \,.
\end{equation*}
Then for any $f \in \ell^\infty(\IZ^n_K)$, we have
\begin{equation*}
\left\| \sum_{k \in \IZ^n_K} f(k)S_{\text{\normalfont per}}^k \right\| 
= \max_{l \in \IZ^n_K} \left| \sum_{k \in \IZ^n_K} f(k) e^{2 \pi i \sum_{j = 1}^n \frac{k_j l_j}{K}} \right|\,.
\end{equation*}
\label{lem:norm_lin_comb_periodic_shifts}
\end{lem}

\begin{proof}
This is readily seen by conjugating the operator $\sum_{k \in \IZ^n_K} f(k)S_{\text{\normalfont per}}^k$ with the discrete Fourier transform,
\begin{equation*}
\phi\mapsto\left(l\mapsto  K^{-\frac{n}{2}} \sum_{m \in \IZ^n_K} \phi(m) e^{-2\pi i \sum_{j = 1}^n \frac{l_j m_j}{K}}\right)\,,
\end{equation*}
yielding the multiplication operator of which the corresponding function is the one within absolute value strokes.
\end{proof}

\begin{prop}{\em (Rieffel's condition)}
For each $f \in \mathcal{S}_{\mathcal{R}}(T^\ast \IT^n)$, we have
\begin{equation*}
\lim_{\hbar \to 0} \|\mathcal{Q}^W_\hbar(f)\| = \|f\|_\infty\,.
\end{equation*}
\label{prop:continuity_of_quantisation_at_0}
\end{prop}

\begin{figure}[H]
\begin{center}
\includegraphics[scale=0.0893]{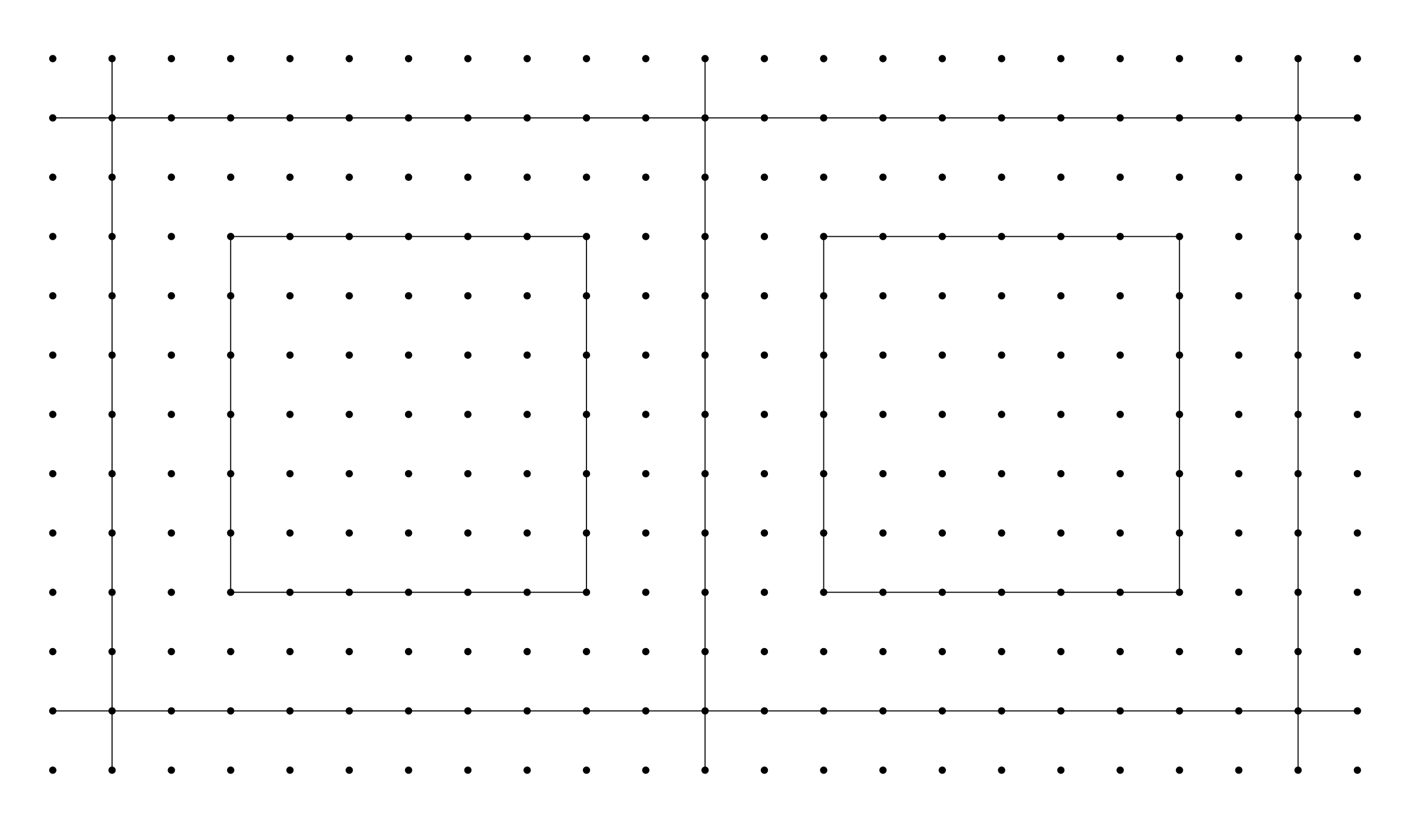}
%\vspace{-0.8cm}
\hspace{-0.3cm}
\captionsetup{width=10cm}
\vspace{-0.065cm}
\caption[width=0.3\linewidth]{Part of the lattice $\Z^2$ with two larger boxes that are adjacent, each of which contains a smaller box.}
\label{fig:boxes}
\end{center}
\end{figure}

\noindent 
Before we give a precise proof of this proposition, it is instructive to first give a sketch of the underlying idea.
To relate the norm of $\mathcal{Q}^W_\hbar(f)$ to that of $f$, we conjugate the quantised function with the Fourier transform to obtain an operator on $\ell^2(\IZ^n)$.
We visualise $\IZ^n$ as a lattice of points in $\IR^n$, and divide it into identical boxes.
In each of these boxes, we identify a slightly smaller box such that all of the smaller boxes are translates of each other in the same way that the larger boxes that contain them are translates of each other. See Figure \ref{fig:boxes}.
\noindent
The difference between the sizes of the small boxes and the sizes of the larger boxes is determined by the values of the various $k_j$ that appear in the function
\begin{equation*}
f = \sum_{j = 1}^m e_{k_j} \otimes h_{U_j, \xi_j, g_j}\,,
\end{equation*}
of which we consider the quantisation; specifically, the shift $S_{k_j} \in B(\ell^2(\IZ^n))$ should always map elements on $\ell^2(\IZ^n)$ supported on points inside the smaller box to functions supported on points inside the larger box containing the small one.
The size of the larger box is determined by a chosen value of $\varepsilon > 0$ and a crude estimate of $\|\mathcal{Q}^W_\hbar(f)\|$.

Given a function $\phi \in \ell^2(\IZ^n)$, we can now estimate the norm of its image under the conjugated quantised function as follows.
First, we consider the projection of $\phi$ onto the subspace of $\ell^2(\IZ^n)$ of elements supported on the set of points inside one of the smaller boxes, and  use the fact that its image under the operator consists of elements supported on the set of points inside the larger box.
We can then consider a periodic version of the operator, and use the preceding lemma to get an estimate on its norm and relate it to the norm of $f$.
Finally, we sum the contributions of all projections of $\phi$ onto the subspaces corresponding to the smaller boxes to obtain an estimate on the difference of the norm of $f$ and that of the conjugated version of its quantisation.
To control the difference between $\phi$ and its projection onto the space corresponding to the union of all of the smaller boxes, we note that the partition into boxes can always be offset by some element of $\IZ^n$ in such a way that the part of $\phi$ supported on the complement of this union is small.

\begin{proof}
Fix $f \in \mathcal{S}_{\mathcal{R}}(T^\ast \IT^n)$ and $\varepsilon > 0$.
We first prove the following statement:
\begin{claimenum}{1}
\item There exists an $\hbar_1 \in (0,\infty)$ such that for each $\hbar \in (0,\hbar_1]$, we have 
\begin{equation*}
\|\mathcal{Q}^W_\hbar(f)\| < \|f\|_\infty + \varepsilon\,.
\end{equation*}
\end{claimenum}

\noindent
%Since $f \in \mathcal{S}_{\mathcal{R}}(T^\ast \IT^n)$, there exist functions $f_1,\ldots, f_m$, where
%\begin{equation*}
%f_j = e_{k_j} \otimes h_{U_j,\xi_j,g_j}\,,
%\end{equation*}
%is a generator of $\mathcal{S}_{\mathcal{R}}(T^\ast \IT^n)$ for $j = 1,\ldots,m$, and $f = \sum_{j = 1}^m f_j$.
%For each $j$, fix the corresponding $k_j,U_j,\xi_j$ and $g_j$ and let
%\begin{equation*}
%h_j := h_{U_j,\xi_j,g_j}\,.
%\end{equation*}
Write $f\in\Sr(T^*\T^n)$ as $f=\sum_{j=1}^m f_j$, where
\begin{equation*}
	f_j = e_{k_j} \otimes h_j\,,
\end{equation*}
for $k_j\in\Z^n$ and $h_j=h_{U_j,\xi_j,g_j}$ for some $U_j,\xi_j$ and $g_j$ (which are not needed in the proof).
Note that by \eqref{eq:formula}, we have a uniform bound on the norms of the operators $(\mathcal{Q}^W_\hbar(f))_{\hbar > 0}$, namely
\begin{equation*}
\|\mathcal{Q}^W_\hbar(f)\| \leq \sum_{j = 1}^m \|h_j\|_\infty = \sum_{j = 1}^m \|g_j\|_\infty =: C\,.
\end{equation*}
Since the case $C = 0$ is trivial, we assume that $C > 0$ (which also implies that $m > 0$).
Now define $L:=\max_{1\leq j\leq m}\supnorm{k_j}$ and fix $K \in\IN \backslash \{0\}$ such that $K \geq 2L$ and such that
\begin{equation}
\left(1 - \frac{2L}{K} \right)^n
> 1 - \left( \frac{\varepsilon}{4C} \right)^2\,.
\label{eq:truncation_error1}
\end{equation}
Moreover, for $j = 1,\ldots,m$, the function $h_j$ is uniformly continuous, hence there exists $\hbar_1 \in (0,\infty)$ such that for each $\hbar \in (0,\hbar_1]$, each $a \in \IZ^n$ and each $b \in \IZ^n$ with $|b_l| < K$ for $l = 1,\ldots,n$, we have
\begin{equation}
|h_j(2 \pi \hbar a) - h_j(\pi \hbar(k_j + 2(a + b)))|
< \frac{\varepsilon}{4m}\,.
\label{eq:periodic_to_full1}
\end{equation}
Now fix $\hbar \in (0,\hbar_1]$, fix $\psi \in L^2(\IT^n)$ with $\|\psi\| = 1$, and let $\phi$ be the image of $\psi$ under the Fourier transform $\mathcal{F} \colon L^2(\IT^n) \rightarrow \ell^2(\IZ^n)$, which we already defined in part (1) of the proof of Proposition \ref{prop:quantisation_notable_properties}.
Furthermore, we define the set
\begin{equation*}
X := \set{a \in \IZ^n}{\text{\normalfont $L \leq a_l < K - L$ for  $l = 1,\ldots,n$}}\,,
\end{equation*}
and we define $K\IZ^n$ and $\IZ^n_K$ as in the previous lemma.
Then, we have
%\begin{align*}
%&\sum_{b + K\IZ^n \in \IZ^n_K} \sum_{a \in X + K\IZ^n} |\phi(a + b)|^2 \\
%&\quad = \sum_{b + K\IZ^n \in \IZ^n_K} \sum_{a \in X} \sum_{a^\prime \in K\IZ^n} |\phi(a + a^\prime + b)|^2 \\
%&\quad = \sum_{a \in X} \sum_{b + K\IZ^n \in \IZ^n_K} \sum_{a^\prime \in K\IZ^n} |\phi(a + a^\prime + b)|^2 \\
%&\quad = \sum_{a \in X} \sum_{b \in \IZ^n} |\phi(a + b)|^2
%= |X| \cdot \sum_{b \in \IZ^n} |\phi(b)|^2
%= |X|\,,
%\end{align*}
\begin{align*}
\sum_{b + K\IZ^n \in \IZ^n_K} \sum_{a \in X + K\IZ^n} |\phi(a + b)|^2
& = \sum_{b + K\IZ^n \in \IZ^n_K} \sum_{a \in X} \sum_{a^\prime \in K\IZ^n} |\phi(a + a^\prime + b)|^2 \\
& = \sum_{a \in X} \sum_{b + K\IZ^n \in \IZ^n_K} \sum_{a^\prime \in K\IZ^n} |\phi(a + a^\prime + b)|^2 \\
& = \sum_{a \in X} \sum_{b \in \IZ^n} |\phi(a + b)|^2
= |X| \cdot \sum_{b \in \IZ^n} |\phi(b)|^2
= |X|\,,
\end{align*}
where
\begin{equation*}
|X| = (K - 2L)^n\,,
\end{equation*}
is the cardinality of the set $X$.
It follows that there exists a $b \in \IZ^n$ with $0 \leq b_l < K$ for $l = 1,\ldots,n$ such that
\begin{equation*}
\sum_{a \in X + K\IZ^n} |\phi(a + b)|^2
\geq |\IZ^n_K|^{-1} (K - 2L)^n
= \left(1 - \frac{2L}{K} \right)^n
> 1 - \left( \frac{\varepsilon}{4C} \right)^2\,.
\end{equation*}
Let $P_{X,b}$ be the orthogonal projection of $\ell^2(\IZ^n)$ onto the subspace
\begin{equation*}
\set{\phi^\prime \in \ell^2(\IZ^n)}{\supp(\phi^\prime) \subseteq b + X + K\IZ^n}\,,
\end{equation*}
so that by the above inequality, we have
\begin{equation}
\begin{aligned}
\|\mathcal{Q}^W_\hbar(f) \mathcal{F}^{-1} (1 - P_{X,b}) \mathcal{F}\psi\|
&\leq \|\mathcal{Q}^W_\hbar(f)\| \|\mathcal{F}^{-1} (1 - P_{X,b}) \phi\| \\
&\leq C \|(1 - P_{X,b})\phi\| \\
&= C \left( 1 - \|P_{X,b} \phi\|^2 \right)^{\frac{1}{2}}
< \frac{\varepsilon}{4}\,.
\end{aligned}
\label{eq:truncation_error2}
\end{equation}
For each $a \in K\IZ^n$, let
\begin{equation*}
P_{a,b} \colon \ell^2(\IZ^n) \rightarrow \ell^2(\IZ^n_K), \quad 
\phi^\prime \mapsto (\,a^\prime + K\IZ^n \mapsto \phi^\prime(a + a^\prime + b) \,)\,,
\end{equation*}
where the representative $a^\prime \in \IZ^n$ has been chosen so that $0 \leq a^\prime_l < K$ for $l = 1,\ldots,n$.
Furthermore, for each $a \in \IZ^n$, we have a corresponding shift operator
\begin{equation*}
S^a \colon \ell^2(\IZ^n) \rightarrow \ell^2(\IZ^n), \quad 
\phi^\prime \mapsto (\,a^\prime \mapsto \phi^\prime(-a + a^\prime)\,)\,,
\end{equation*}
and for each $a + K\IZ^n \in \IZ^n_K$, we define the shift operator $S_{\text{\normalfont per}}^{a + K\IZ^n}$ as in the previous lemma.
Finally, for each $a \in K\IZ^n$, we define
\begin{equation*}
A_{a,b} := \sum_{j = 1}^m h_j(2 \pi \hbar(a + b)) S_{\text{\normalfont per}}^{k_j + K\IZ^n}\,.
\end{equation*}
Using Lemma \ref{lem:norm_lin_comb_periodic_shifts}, we obtain
\begin{equation}
\begin{aligned}
\|A_{a,b}\|
&= \max_{a^\prime + K\IZ^n \in \IZ^n_K} \left| \sum_{j = 1}^m e^{2 \pi i \frac{k_j\cdot a^\prime}{K}} h_j(2 \pi \hbar(a + b)) \right| \\
&\leq \sup_{[x] \in \IT^n} \left| \sum_{j = 1}^m e^{2 \pi i k_j \cdot x} h_j(2 \pi \hbar(a + b)) \right| \\
&\leq \sup_{([x],p) \in \IT^n \times \IR^n} \left| \sum_{j = 1}^m e^{2 \pi i k_j \cdot x} h_j(p) \right| \\
&= \|f\|_\infty\,.
\end{aligned}
\label{eq:norm_of_periodic_operator}
\end{equation}
Moreover, using our explicit formula \eqref{eq:formula}, we find that
\begin{align*}
&P_{a,b} \mathcal{F} \mathcal{Q}^W_\hbar(f) \mathcal{F}^{-1} P_{X,b} \phi \\
&\quad= P_{a,b} \mathcal{F} \mathcal{Q}^W_\hbar(f) \mathcal{F}^{-1} P_{X,b} \sum_{a^\prime \in \IZ^n} \phi(a^\prime) \delta_{a^\prime} \\
&\quad= P_{a,b} \sum_{a^\prime \in b + X + K\IZ^n} \sum_{j = 1}^m h_j(\pi \hbar(k_j + 2 a^\prime)) \phi(a^\prime) \delta_{a^\prime + k_j} \\
&\quad= \sum_{a^\prime \in X} \sum_{j = 1}^m h_j(\pi \hbar(k_j + 2(a + b + a^\prime))) \phi(a + b + a^\prime) \delta_{a^\prime + k_j + K\IZ^n} \\
&\quad= \sum_{j = 1}^m S_{\text{\normalfont per}}^{k_j + K\IZ^n} \sum_{a^\prime \in X} h_j(\pi \hbar(k_j + 2(a + b + a^\prime))) \phi(a + b + a^\prime) \delta_{a^\prime + K\IZ^n}\,,
\end{align*}
where in the third step, we have used the fact that $a' + k_j \in\{0,\ldots, K-1\}^n$ for each $a^\prime \in X$ and $j = 1,\ldots,m$.
On the other hand, we have
\begin{align*}
A_{a,b} P_{a,b} P_{X,b} \phi
&= A_{a,b} P_{a,b} P_{X,b} \sum_{a^\prime \in \IZ^n} \phi(a^\prime) \delta_{a^\prime}
= A_{a,b} \sum_{a^\prime \in X} \phi(a + b + a^\prime) \delta_{a^\prime + K\IZ^n} \\
&= \sum_{j = 1}^m S_{\text{\normalfont per}}^{k_j + K\IZ^n} \sum_{a^\prime \in X} h_j(2 \pi \hbar(a + b)) \phi(a + b + a^\prime) \delta_{a^\prime + K\IZ^n}\,.
\end{align*}
Writing
\begin{equation*}
\mu_{a^\prime,j}
:= h_j(2 \pi \hbar(a + b)) - h_j(\pi \hbar(k_j + 2(a + b + a^\prime)))\,,
\end{equation*}
for $j = 1,\ldots,m$ and $a^\prime \in X$, we obtain
\begin{equation}
\begin{aligned}
&\|(A_{a,b} P_{a,b} P_{X,b} - P_{a,b} \mathcal{F} \mathcal{Q}^W_\hbar(f) \mathcal{F}^{-1} P_{X,b})\phi\| \\
&\quad= \left\| \sum_{j = 1}^m S_{\text{\normalfont per}}^{k_j + K\IZ^n} \sum_{a^\prime \in X} \mu_{a^\prime,j}  \phi(a + b + a^\prime) \delta_{a^\prime + K\IZ^n} \right\| \\
&\quad\leq \sum_{j = 1}^m \left\| \sum_{a^\prime \in X} \mu_{a^\prime,j}  \phi(a + b + a^\prime) \delta_{a^\prime + K\IZ^n} \right\| \\
&\quad= \sum_{j = 1}^m \left( \sum_{a^\prime \in X} |\mu_{a^\prime,j}|^2 |\phi(a + b + a^\prime)|^2 \right)^{\frac{1}{2}} \\
&\quad\leq m \cdot \max_{a^{\prime \prime} \in X} |\mu_{a^{\prime \prime},j}| \left( \sum_{a^\prime \in X} |\phi(a + b + a^\prime)|^2 \right)^{\frac{1}{2}} \\
&\quad< \frac{\varepsilon}{4} \|P_{a,b} P_{X,b} \phi\|\,,
\end{aligned}
\label{eq:periodic_to_full2}
\end{equation}
where we have used equation \eqref{eq:periodic_to_full1} in the final step.
From equations \eqref{eq:norm_of_periodic_operator} and \eqref{eq:periodic_to_full2}, we obtain
\begin{align*}
&\|P_{a,b} \mathcal{F} \mathcal{Q}^W_\hbar(f) \mathcal{F}^{-1} P_{X,b} \phi\| \\
&\quad \leq \|A_{a,b} P_{a,b} P_{X,b} \phi\| + \|(A_{a,b} P_{a,b} P_{X,b} - P_{a,b} \mathcal{F} \mathcal{Q}^W_\hbar(f) \mathcal{F}^{-1} P_{X,b})\phi\| \\
&\quad < \left( \|f\|_\infty + \frac{\varepsilon}{4} \right) \|P_{a,b} P_{X,b} \phi\|\,,
\end{align*}
%\begin{align*}
%\|P_{a,b} \mathcal{F} \mathcal{Q}^W_\hbar(f) \mathcal{F}^{-1} P_{X,b} \phi\|
%& \leq \|A_{a,b} P_{a,b} P_{X,b} \phi\| + \|(A_{a,b} P_{a,b} P_{X,b} - P_{a,b} \mathcal{F} \mathcal{Q}^W_\hbar(f) \mathcal{F}^{-1} P_{X,b})\phi\| \\
%& < \left( \|f\|_\infty + \frac{\varepsilon}{4} \right) \|P_{a,b} P_{X,b} \phi\|\,,
%\end{align*}
for each $a \in K\IZ^n$.
It is straightforward to see that for each $\phi^\prime \in \ell^2(\IZ^n)$, we have
\begin{equation*}
\sum_{a \in K\IZ^n} \|P_{a,b} \phi^\prime \|^2 = \|\phi^\prime\|^2\,,
\end{equation*}
so
\begin{align*}
\|\mathcal{Q}^W_\hbar(f) \mathcal{F}^{-1} P_{X,b} \phi\|^2
&= \|\mathcal{F} \mathcal{Q}^W_\hbar(f) \mathcal{F}^{-1} P_{X,b} \phi\|^2
= \sum_{a \in K\IZ^n} \|P_{a,b} \mathcal{F} \mathcal{Q}^W_\hbar(f) \mathcal{F}^{-1} P_{X,b} \phi\|^2 \\
&< \sum_{a \in K\IZ^n} \left( \|f\|_\infty + \frac{\varepsilon}{4} \right)^2 \|P_{a,b} P_{X,b} \phi\|^2 \\
&= \left( \|f\|_\infty + \frac{\varepsilon}{4} \right)^2 \|P_{X,b} \phi\|^2
\leq \left( \|f\|_\infty + \frac{\varepsilon}{4} \right)^2\,,
\end{align*}
which together with equation \eqref{eq:truncation_error2} implies
\begin{align*}
\|\mathcal{Q}^W_\hbar(f) \psi\|
&\leq \|\mathcal{Q}^W_\hbar(f) \mathcal{F}^{-1} P_{X,b} \mathcal{F} \psi\| + \|\mathcal{Q}^W_\hbar(f) \mathcal{F}^{-1} (1 - P_{X,b}) \mathcal{F} \psi\| \\
&< \|f\|_\infty + \frac{\varepsilon}{4} + \frac{\varepsilon}{4}
= \|f\|_\infty + \frac{\varepsilon}{2}\,,
\end{align*}
and since $\psi \in L^2(\IT^n)$ was an arbitrary vector with norm $1$, we obtain
\begin{equation*}
\|\mathcal{Q}^W_\hbar(f)\|
\leq \|f\|_\infty + \frac{\varepsilon}{2}
< \|f\|_\infty + \varepsilon\,,
\end{equation*}
for each $\hbar \in (0,\hbar_1]$ which proves (a).\\

\noindent
We now turn to the reverse inequality:
\begin{claimenum}{2}
\item There exists an $\hbar_2 \in (0,\infty)$ such that for each $\hbar \in (0,\hbar_2]$, we have
\begin{equation*}
\|f\|_\infty < \|\mathcal{Q}^W_\hbar(f)\| + \varepsilon\,.
\end{equation*}
\end{claimenum}

\noindent
Let $(x,p) \in [0,1)^n \times \IR^n$ be a point such that
\begin{equation*}
\|f\|_\infty 
< |f([x],p)| + \frac{\varepsilon}{8}\,.
\end{equation*}
By uniform continuity of $f$, there exists a $\delta > 0$ such that for each $(x',p') \in (-1,1)^n \times \IR^n$ with $\sum_{l = 1}^n |x'_l - x_{l}| + |p'_l - p_{l}| < \delta$, we have
\begin{equation*}
|f([x],p) - f([x'],p')| < \frac{\varepsilon}{8}\,.
\end{equation*}
Now fix $L \in \IN$ as in the proof of part (a), and fix $K \in \IN\backslash\{0\}$ in such a way that equation \eqref{eq:truncation_error1} holds, and that we have
\begin{equation}
K > \max \left( 2L, \frac{2n}{\delta} \right)\,.
\label{eq:sufficiently_small_stepsize1}
\end{equation}
Furthermore, fix $\hbar_2 > 0$ such that equation \eqref{eq:periodic_to_full1} holds for each $\hbar \in (0,\hbar_2]$, and that we have
\begin{equation}
2\pi \hbar_2 K < \frac{\delta}{2n}\,.
\label{eq:sufficiently_small_hbar_for_momentum1}
\end{equation}
Now fix such an $\hbar \in (0,\hbar_2]$.
Next, we note that by equation \eqref{eq:sufficiently_small_hbar_for_momentum1} there exists an $a \in K\IZ^n$ such that
\begin{equation*}
p_{l} - \frac{\delta}{2n}
< 2 \pi \hbar a_l
\leq p_{l}\,,
\end{equation*}
and that by equation \eqref{eq:sufficiently_small_stepsize1}, there exists a $b \in \{0,\ldots,K-1\}^n$ such that
\begin{equation*}
\left|\frac{b_l}{K} - x_{l}\right| < \frac{\delta}{2n}\,,
\end{equation*}
for $l = 1,\ldots,n$.
Fix such $a$ and $b$.
It follows that
\begin{equation*}
\sum_{l = 1}^n \left|\frac{b_l}{K} - x_{l}\right| + |2 \pi \hbar a_l - p_{l}| < \delta\,,
\end{equation*}
so that
\begin{equation*}
\left|\sum_{j = 1}^m e^{2\pi i \frac{k_j\cdot b}{K}} h_j(2 \pi \hbar a) - f([x], p) \right|
< \frac{\varepsilon}{8}\,,
\end{equation*}
and therefore, by the triangle inequality and our choice of $([x] , p)$,
\begin{equation*}
\left| \left| \sum_{j = 1}^m e^{2\pi i \frac{k_j\cdot b}{K}} h_j(2 \pi \hbar a) \right| - \|f\|_\infty \right|
<  \frac{\varepsilon}{4}\,.
\end{equation*}
Now define $\phi \in \ell^2(\IZ^n)$ by
\begin{equation*}
\phi(a^\prime) :=
\left\{
\begin{array}{l l}
\displaystyle K^{-\frac{n}{2}} e^{-2 \pi i \frac{a^\prime\cdot b}{K}} & \quad \text{\normalfont if $0 \leq a^\prime_l - a_l < K$ for $l = 1,\ldots,n$,} \\
0 & \quad \text{\normalfont otherwise,}
\end{array}
\right.
\end{equation*}
and let $\psi := \mathcal{F}^{-1}\phi \in L^2(\IT^n)$.
Then $\|\psi\| = \|\phi\| = 1$, and
\begin{equation*}
A_{a,0}P_{a,0}\phi 
= \sum_{j = 1}^m e^{2\pi i \frac{k_j\cdot b}{K}} h_j(2 \pi \hbar a) P_{a,0}\phi\,,
\end{equation*}
with $A_{a,b}$ and $P_{a,b}$ as defined in part (a). Since $\norm{P_{a,0}\phi}=1$, we have
\begin{equation*}
\|A_{a,0}P_{a,0}\phi\|
= \left| \sum_{j = 1}^m e^{2\pi i \frac{k_j \cdot b}{K}} h_j(2 \pi \hbar a) \right|
> \|f\|_\infty - \frac{\varepsilon}{4}\,.
\end{equation*}
Defining $X$ in the same way as we did in the proof part (a), it follows that
\begin{align*}
\|A_{a,0}P_{a,0}P_{X,0}\phi\|
&\geq \|A_{a,0}P_{a,0}\phi\| - \|A_{a,0}\| \|(1 - P_{X,0})\phi\| \\
&> \|f\|_\infty - \frac{\varepsilon}{4} - \frac{\varepsilon}{4}
= \|f\|_\infty - \frac{\varepsilon}{2}\,.
\end{align*}
Next, we note that the function $\mathcal{F} \mathcal{Q}^W_\hbar(f) \mathcal{F}^{-1} P_{X,0} \phi:\Z^n\rightarrow\C$ is supported in the set of $a^\prime \in \IZ^n$ satisfying $a_l \leq a^\prime_l < a_l + K$ for $l = 1,\ldots,n$.
Combining this observation with the estimate just obtained and equation \eqref{eq:periodic_to_full2} yields
\begin{align*}
\|\mathcal{F} \mathcal{Q}^W_\hbar(f) \mathcal{F}^{-1} P_{X,0} \phi\|
&= \|P_{a,0} \mathcal{F} \mathcal{Q}^W_\hbar(f) \mathcal{F}^{-1} P_{X,0} \phi\| \\
&\geq \|A_{a,0}P_{a,0}P_{X,0}\phi\| - \|(A_{a,0} P_{a,0} P_{X,0} - P_{a,0} \mathcal{F} \mathcal{Q}^W_\hbar(f) \mathcal{F}^{-1} P_{X,0})\phi\| \\
&> \|f\|_\infty - \frac{\varepsilon}{2} - \frac{\varepsilon}{4}
= \|f\|_\infty - \frac{3\varepsilon}{4}\,.
\end{align*}
We use this together with equation \eqref{eq:truncation_error2} to obtain
\begin{align*}
\|\mathcal{Q}^W_\hbar(f)\psi\|
&= \|\mathcal{F} \mathcal{Q}^W_\hbar(f) \psi\| \\
&\geq \|\mathcal{F} \mathcal{Q}^W_\hbar(f)\mathcal{F}^{-1}P_{X,0}\phi\| - \|\mathcal{Q}^W_\hbar(f) \mathcal{F}^{-1} (1 - P_{X,0}) \mathcal{F}\psi\| \\
&> \|f\|_\infty - \frac{3\varepsilon}{4} - \frac{\varepsilon}{4}
= \|f\|_\infty - \varepsilon\,.
\end{align*}
Since $\|\psi\| = 1$, this establishes (b).\\

\noindent 
Finishing up the proof, taking $\hbar_0 := \min(\hbar_1,\hbar_2)$, we infer that for each $\hbar \in (0,\hbar_0]$, we have $|\|\mathcal{Q}^W_\hbar(f)\| - \|f\|_\infty| < \varepsilon$, hence $\lim_{\hbar \to 0} \|\mathcal{Q}^W_\hbar(f)\| = \|f\|_\infty$, as desired.
\end{proof}

\noindent 
We are now ready to prove the main result of this subsection.
Let $\mathcal{Q}^W_0 := \Id_{\mathcal{S}_{\mathcal{R}}(T^\ast \IT^n)}$, let $A_0$ be the C$^\ast$-algebra $C_{\mathcal{R}}(T^\ast \IT^n)$.
In the following theorem, it should be understood that $\|\mathcal{Q}^W_\hbar(f)\| := \|f\|_\infty$ for $\hbar = 0$.

\begin{thrm}
Let $I \subset [0,\infty)$ be a subset containing $0$ as an accumulation point.
Then, except for continuity at $\hbar > 0$, the triple
\begin{equation*}
(I, (A_\hbar)_{\hbar \in I}, (\mathcal{Q}^W_\hbar \colon \mathcal{S}_{\mathcal{R}}(T^\ast \IT^n) \rightarrow A_\hbar)_{\hbar \in I}) \,,
\end{equation*}
is a strict quantisation of the Poisson algebra $\mathcal{S}_{\mathcal{R}}(T^\ast \IT^n)$, i.e., it satisfies
\begin{myenum}
\item Rieffel's condition at $\hbar=0$: for each $f \in \mathcal{S}_{\mathcal{R}}(T^\ast \IT^n)$, the function $\hbar \mapsto \|\mathcal{Q}^W_\hbar(f)\|$ is continuous at $0$.

\item Von Neumann's condition: for each $f, g \in \mathcal{S}_{\mathcal{R}}(T^\ast \IT^n)$, we have
\begin{equation*}
\lim_{\substack{\hbar \to 0\\ \hbar \in I}} \|\mathcal{Q}^W_\hbar(f) \mathcal{Q}^W_\hbar(g) - \mathcal{Q}^W_\hbar(fg)\| = 0\,.
\end{equation*}

\item Dirac's condition: for each $f, g \in \mathcal{S}_{\mathcal{R}}(T^\ast \IT^n)$, we have
\begin{equation*}
\lim_{\substack{\hbar \to 0\\ \hbar \in I}} \|(-i\hbar)^{-1}[\mathcal{Q}^W_\hbar(f),\mathcal{Q}^W_\hbar(g)] - \mathcal{Q}^W_\hbar(\{f,g\})\| = 0\,.
\end{equation*}

\item Completeness: for each $\hbar \in I$, the set $\mathcal{Q}^W_\hbar(\mathcal{S}_{\mathcal{R}}(T^\ast \IT^n))$ is dense in $A_\hbar$.
\end{myenum}
\label{thrm:Weyl_quantisation_is_strict}
\end{thrm}

\begin{proofenum}
\item This was shown in Proposition \ref{prop:continuity_of_quantisation_at_0}.

\item First suppose that $f_j$ is a generator $e_{k_j} \otimes h_{U_j,\xi_j,g_j}$ of $\mathcal{S}_{\mathcal{R}}(T^\ast \IT^n)$ for $j = 1,2$.
As in Proposition \ref{prop:continuity_of_quantisation_at_0}, we will write $h_j$ instead of $h_{U_j,\xi_j,g_j}$.
Let $k := k_1 + k_2$.
Then
\begin{equation*}
f_1 \cdot f_2
= (e_{k_1} \otimes h_1) \cdot (e_{k_2} \otimes h_2)
= e_k \otimes (h_1 \cdot h_2)\,.
\end{equation*}
Applying part (3) of Proposition \ref{prop:Weyl_quantisation_is_well-defined} yields
\begin{equation*}
\mathcal{Q}^W_\hbar(f_1 f_2)\psi_a 
= (h_1 \cdot h_2)(\pi \hbar(k + 2a)) \psi_{k + a}\,.
\end{equation*}
for each $a \in \IZ^n$.
On the other hand, we have
\begin{equation}
\begin{aligned}
\mathcal{Q}^W_\hbar(f_1) \mathcal{Q}^W_\hbar(f_2) \psi_a
&= h_2(\pi \hbar(k_2 + 2a)) \mathcal{Q}^W_\hbar(f_1) \psi_{k_2 + a} \\
&= h_1(\pi \hbar(k_1 + 2(k_2 + a))) h_2(\pi \hbar(k_2 + 2a)) \psi_{k + a} \\
&= h_1(\pi \hbar(k + k_2 + 2a)) \cdot h_2(\pi \hbar(k - k_1 + 2a)) \cdot \psi_{k + a}\,,
\end{aligned}
\label{eq:product_of_quantisations}
\end{equation}
so
\begin{equation*}
\begin{aligned}
(\mathcal{Q}^W_\hbar(f_1) \mathcal{Q}^W_\hbar(f_2) - \mathcal{Q}^W_\hbar(f_1 f_2))\psi_a 
& = \left( h_1 (\pi \hbar (k + k_2 + 2a)) \cdot h_2(\pi \hbar (k - k_1 + 2a)) \right)\\
&\quad \left. - (h_1 \cdot h_2)(\pi \hbar (k + 2a)) \right) \psi_{k + a}\,,
\end{aligned}
\end{equation*}
for each $a \in \IZ^n$.
Now let $c_{a,\hbar}^{(1)}$ be the scalar in front of $\psi_{k + a}$ on the right-hand side of the last equation.
It is not hard to see from this equation that
\begin{equation*}
\|\mathcal{Q}^W_\hbar(f_1) \mathcal{Q}^W_\hbar(f_2) - \mathcal{Q}^W_\hbar(f_1 f_2)\|
\leq \sup_{a \in \IZ^n} |c_{a,\hbar}^{(1)}|\,,
\end{equation*}
for each $\hbar > 0$.
Now note for $j = 1,2$, all derivatives of $h_j\in\Wr(\R^n)$ are bounded, and so $h_j$ is Lipschitz continuous. This implies that
\begin{align*}
	h_1(\pi\hbar(k+k_2+2a))&=h_1(\pi\hbar(k+2a))+\O(\hbar) \,,\\
	h_2(\pi\hbar(k-k_1+2a))&=h_2(\pi\hbar(k+2a))+\O(\hbar) \,,
\end{align*}
where big O notation signifies a limit of $\hbar\rightarrow0$, \emph{uniformly in $a$}, analogous to the notation in \textsection\ref{sct:trigonometric potentials}. When plugging the above formulas into the definition of $c_{a,\hbar}^{(1)}$ and using the fact that $h_1$ and $h_2$ are bounded functions, we find that
	$$c_{a,\hbar}^{(1)}=\O(\hbar)\,,$$
and therefore
\begin{equation*}
\lim_{\substack{\hbar \to 0\\ \hbar \in I}} \|\mathcal{Q}^W_\hbar(f_1) \mathcal{Q}^W_\hbar(f_2) - \mathcal{Q}^W_\hbar(f_1 f_2)\| = 0\,.
\end{equation*}
By bilinearity, this result extends to arbitrary $f_1, f_2 \in \mathcal{S}_{\mathcal{R}}(T^\ast \IT^n)$.

\item As in the previous part of the proof, we prove the statement for $f_j = e_{k_j} \otimes h_j$, from which the general case readily follows.
We have
\begin{align*}
\{f_1, f_2\} 
& = \sum_{l = 1}^n \left(\frac{\partial f_1}{\partial p_l} \frac{\partial f_2}{\partial q_l} - \frac{\partial f_1}{\partial q_l} \frac{\partial f_2}{\partial p_l}\right) \\
& = \sum_{l = 1}^n \left( e_{k_1} \otimes \frac{\partial h_1}{\partial p_l} \right) \cdot \left( \frac{\partial e_{k_2}}{\partial q_l} \otimes h_2 \right) - \left( \frac{\partial e_{k_1}}{\partial q_l} \otimes h_1 \right) \cdot \left( e_{k_2} \otimes \frac{\partial h_2}{\partial p_l} \right) \\
& = 2\pi i e_k \otimes \left( (\nabla_{k_2}h_1) h_2 - h_1 \nabla_{k_1}h_2 \right)\,,
\end{align*}
where $k = k_1 + k_2$, as in part (2) of this theorem, and $\nabla_{v}h$ is the directional derivative of $h$ in the direction of $v$.
Applying part (3) of Proposition \eqref{prop:Weyl_quantisation_is_well-defined} yields
\begin{equation*}
\mathcal{Q}^W_\hbar(\{f_1, f_2\})\psi_a
= 2\pi i \left( (\nabla_{k_2}h_1) h_2 - h_1 \nabla_{k_1}h_2 \right)(\pi \hbar(k + 2a)) \psi_{k + a}\,,
\end{equation*}
while equation \eqref{eq:product_of_quantisations} yields
\begin{align*}
[\mathcal{Q}^W_\hbar(f_1),\mathcal{Q}^W_\hbar(f_2)]\psi_a 
&= \left( h_1(\pi \hbar(k + k_2 + 2a)) \cdot h_2(\pi \hbar(k - k_1 + 2a)) \right. \\
&\left. \quad {} - h_1(\pi \hbar(k - k_2 + 2a)) \cdot h_2(\pi \hbar(k + k_1 + 2a)) \right) \psi_{k + a}\,.
\end{align*}
It follows that
\begin{equation*}
\left( (-i\hbar)^{-1}[\mathcal{Q}^W_\hbar(f_1),\mathcal{Q}^W_\hbar(f_2)] - \mathcal{Q}^W_\hbar(\{f_1, f_2\}) \right)\psi_a
= c_{a,\hbar}^{(2)} \psi_{k + a}\,,
\end{equation*}
where for each $a \in \IZ^n$ and each $\hbar > 0$, we define
\begin{align*}
c_{a,\hbar}^{(2)}
:=& (-i\hbar)^{-1} \left( h_1(\pi \hbar(k + k_2 + 2a)) \cdot h_2(\pi \hbar(k - k_1 + 2a)) \right. \\*
& \hphantom{(-i\hbar)^{-1} \left( \right.} \left. - h_1(\pi \hbar(k - k_2 + 2a)) \cdot h_2(\pi \hbar(k + k_1 + 2a)) \right)\\*
&-2\pi i \left( (\nabla_{k_2} h_1) h_2 - h_1  \nabla_{k_1} h_2 \right)(\pi \hbar(k + 2a))\,.
\end{align*}
It is readily seen that
\begin{equation*}
\left\|(-i\hbar)^{-1}[\mathcal{Q}^W_\hbar(f_1),\mathcal{Q}^W_\hbar(f_2)] - \mathcal{Q}^W_\hbar(\{f_1, f_2\}) \right\|
\leq \sup_{a \in \IZ^n} |c_{a,\hbar}^{(2)}|\,.
\end{equation*}
We claim that the right-hand side of this inequality converges to $0$ as $\hbar \in I$ goes to 0; evidently, this will show that Dirac's condition holds.

Because the second order derivatives of $h_j$ are bounded, Taylor's theorem gives
\begin{equation}
	h_j(\pi \hbar(k + v + 2a)) - h_j(\pi\hbar(k + 2a)) - \pi \hbar \nabla_v h_j(\pi\hbar(k + 2a)) =\O(\hbar^2) \,,
\label{eq:Taylor's_theorem_estimate}
\end{equation}
for each $v \in \IR^n$ and $j = 1,2$.
Dividing the expression on the left-hand side of \eqref{eq:Taylor's_theorem_estimate} by $-i\hbar$ yields
\begin{align*}
&(-i\hbar)^{-1} (h_j(\pi \hbar(k + v + 2a)) - h_j(\pi\hbar(k + 2a))) - \pi i \nabla_v h_j(\pi\hbar(k + 2a))=\O(\hbar)\,.
\end{align*}
This can be used to show that $c_{a,\hbar}^{(2)} \to 0$ uniformly in $a \in \IZ^n$ as $\hbar \to 0$, which proves the claim.

\item According to part (2) of Proposition \ref{prop:Poisson_subalg_for_arbitrary_n}, the space $\mathcal{S}_{\mathcal{R}}(T^\ast \IT^n)$ is a $^\ast$-subalgebra of $C_{\mathcal{R}}(T^\ast \IT^n)$.
According to part (1) of Proposition \ref{prop:quantisation_notable_properties} the Weyl quantisation map is linear and compatible with the involutions on the algebras involved.
Moreover, it is readily seen from our computation of $\mathcal{Q}^W_\hbar(f_1)\mathcal{Q}^W_\hbar(f_2)$ in the proof of part (2) of this theorem that $\mathcal{Q}^W_\hbar(\mathcal{S}_{\mathcal{R}}(T^\ast \IT^n))$ is closed under multiplication.
Thus $\mathcal{Q}^W_\hbar(\mathcal{S}_{\mathcal{R}}(T^\ast \IT^n))$ is a $^\ast$-algebra.
It follows that $A_\hbar$, which is by definition the smallest C$^\ast$-algebra that contains $\mathcal{Q}^W_\hbar(\mathcal{S}_{\mathcal{R}}(T^\ast \IT^n))$, is the closure of $\mathcal{Q}^W_\hbar(\mathcal{S}_{\mathcal{R}}(T^\ast \IT^n))$.
\end{proofenum}

\begin{rem}
The statement that for arbitrary $f \in \mathcal{S}_{\mathcal{R}}(T^\ast \IT^n)$, the map
\begin{equation*}
[0,\infty) \rightarrow [0,\infty) \,, \quad 
\hbar \mapsto \|\mathcal{Q}^W_\hbar(f)\|\,,
\end{equation*}
is continuous at points other than $\hbar = 0$ is false.
As a counterexample, let $\hbar_0 > 0$ be arbitrary, and consider the function $f = e_0 \otimes h$, where the function $h$ is defined as follows:
\begin{equation*}
h \colon \IR^n \rightarrow \IR, \quad 
p = (p_1,p_2,\ldots, p_n) \mapsto \sin \left( \frac{p_1}{\hbar_0} \right)\,.
\end{equation*}
Note that $h$ can be written as the sum of two generators of $\Weylalg{\IR^n} \subseteq \Teunalg{\IR^n}$, so $f \in \mathcal{S}_{\mathcal{R}}(T^\ast \IT^n)$.
Futhermore, $h$ vanishes at each point in $2 \pi \hbar_0 \cdot \IZ^n$, hence $\mathcal{Q}^W_{\hbar_0}(f) = 0$ by the explicit formula \eqref{eq:formula}, or equivalently, $\|\mathcal{Q}^W_{\hbar_0}(f)\| = 0$.
On the other hand, for each $N \in \IN \backslash \{0\}$, let
\begin{equation*}
\hbar_N := \hbar_0 \left( 1 + \frac{1}{4N} \right)\,.
\end{equation*}
Then $\|\mathcal{Q}^W_{\hbar_N}(f)\| = 1$; indeed, we have $\|\mathcal{Q}^W_{\hbar_N}(f)\| \leq \|h\|_\infty = 1$, and equality holds since
\begin{equation*}
\mathcal{Q}^W_{\hbar_N}(f) \psi_{(N,0,0,\ldots,0)}
= \psi_{(N,0,0,\ldots,0)}\,.
\end{equation*}
Thus, while $\lim_{N \to \infty} \hbar_N = \hbar_0$, we also have
\begin{equation*}
\lim_{N \to \infty} \|\mathcal{Q}^W_{\hbar_N}(f)\|
= 1 \neq 0
= \|\mathcal{Q}^W_{\hbar_0}(f)\|\,,
\end{equation*}
so the function $\hbar \to \|\mathcal{Q}^W_\hbar(f)\|$ fails to be continuous at $\hbar_0$.

The issue of continuity of the norm of the quantisation of a given function at points $\hbar \neq 0$ is often sidestepped in the literature for reasons related to geometric quantisation, which imposes the condition that $\hbar$ be of the form $\hbar_0/m$, $m \in \IN \backslash \{0\}$ for some fixed $\hbar_0 > 0$
(cf. \cite{hawkins08} for a discussion of this point, and also a nice overview of the various notions of quantisation throughout the literature).
In such cases the set $I \backslash \{0\}$ in the above theorem is a discrete subset of $(0,\infty)$, so the restriction of $\hbar \to \|\mathcal{Q}^W_\hbar(f)\|$ to $I$ is trivially continuous at all points outside of $0$, and the family of quantisation maps constitutes an actual strict quantisation.
\end{rem}

\noindent 
The fact that $\QW$ is not quite a strict quantisation as defined in \cite[Definition II.1.1.1]{landsman98} is most likely a consequence of the fact that the quantum resolvent algebra defined in this paper is by construction already (faithfully) represented on a Hilbert space, namely $L^2(\T^n)$.
Despite the fact that the norm of the quantisation of a function is not continuous for $\hbar > 0$, we still have continuity of quantisation in another way:

\begin{prop}
Let $f \in \mathcal{S}_{\mathcal{R}}(T^\ast \IT^n)$.
Then the map
\begin{equation*}
(0,\infty) \rightarrow A_\hbar \subseteq B(L^2(\IT^n)) \,, \quad 
\hbar \mapsto \mathcal{Q}^W_\hbar(f)\,,
\end{equation*}
is continuous with respect to the strong operator topology on the codomain.
\end{prop}

\begin{proof}
By linearity of the quantisation map and the fact that $\mathcal{S}_{\mathcal{R}}(T^\ast \IT^n)$ is the linear span of generators of $\mathcal{C}_{\mathcal{R}}(T^\ast \IT^n)$, we may assume without loss of generality that there exists a $k \in \IZ^n$ and a generator $h$ of $\Teunalg{\IR^n}$ such that $f = e_k \otimes h$.
%%%%
Then, by our explicit formula \eqref{eq:formula}, we find
	$$\norm{\QW(f)\psi_l-\mathcal{Q}^W_{\hbar_0}(f)\psi_l}=|h(\pi\hbar(k+2l))-h(\pi\hbar_0(k+2l))|\rightarrow0\,,$$
whenever $\hbar\rightarrow\hbar_0$ in $(0,\infty)$. This convergence also holds when we replace $\psi_l$ by a vector in $\spn_l\{\psi_l\}$.
Furthermore, in part (3) of Proposition \ref{prop:Weyl_quantisation_is_well-defined}, we have seen that
	$$\norm{\QW(f)}\leq\supnorm{h}\,.$$
Now let $\psi\in L^2(\T^n)$ be arbitrary.
Fix $\epsilon > 0$.
Since $\spn_l\{\psi_l\}$ is dense in $L^2(\T^n)$, there exists $\tilde{\psi} \in \spn_l\{\psi_l\}$ such that $\|\tilde{\psi} - \psi\| < \varepsilon / (4(\|h\|_\infty + 1))$.
By the discussion above, there exists $\delta > 0$ such that $\norm{\QW(f)\psi-\mathcal{Q}^W_{\hbar_0}(f)\psi} < \epsilon / 2$ whenever $\hbar > 0$ satisfies $|\hbar - \hbar_0| < \delta$.
Then for any such $\hbar$, we have
\begin{align*}
&\norm{\QW(f)\psi-\mathcal{Q}^W_{\hbar_0}(f)\psi}\\
&\quad \leq\norm{\QW(f)(\psi-\tilde{\psi})}
+ \norm{\QW(f)\tilde{\psi}-\mathcal{Q}^W_{\hbar_0}(f)\tilde{\psi}}
+ \norm{\mathcal{Q}^W_{\hbar_0}(f)(\tilde{\psi}-\psi)} \\
&\quad \leq 2\supnorm{h}\norm{\tilde{\psi}-\psi} 
+ \norm{\QW(f)\tilde{\psi}-\mathcal{Q}^W_{\hbar_0}(f)\tilde{\psi}}\,
< \varepsilon \,,
\end{align*} 
which concludes the proof of the proposition.
\end{proof}

\ifx\mycmd\undefined
	\phantomsection
	\addcontentsline{toc}{section}{References}
	\bibliographystyle{abbrv}
	\bibliography{./../Miscellaneous/References}

	\end{document}
\fi

\ifx\mycmd\undefined
	\documentclass[a4paper, 11pt, leqno]{article}
	
	\begin{document}
\fi

\section{Quantum time evolution}
\label{sec:quantum time evolution}
\noindent 
Our next task is to show that $A_\hbar=C^*\!\left(\QW(\Sr(T^*\T^n))\right)$ is invariant under time evolution for each Hamiltonian with potential $V \in C(\T^n)$.
The general proof strategy resembles that of Buchholz and Grundling in \cite[Proposition 6.1]{buchholz08}.
However, the present setting differs from theirs in two important ways, each of which introduces its own technical problems.
First of all, our configuration space is $\T^n$ rather than $\R^n$.
Secondly, we consider the problem of invariance under time evolution for arbitrary $n \in \N$, whereas Buchholz and Grundling only discuss the case $n = 1$.
We start with the simplest type of time evolution:

\begin{lem}\label{lem:free quantum time evolution}
Let $\hbar > 0$.
The algebra $A_\hbar$ is closed under the quantum time evolution corresponding to the free Hamiltonian $H_0$ that is the unique self-adjoint extension of the essentially self-adjoint operator $-\frac{\hbar^2}{2} \sum\frac{d^2}{dx_j^2}$ with domain $C^\infty(\IT^n)$.
\end{lem}

\begin{rem}
The fact that for any compact Riemannian manifold $M$ the Laplace--Beltrami operator on $C^\infty(M)$ has a unique self-adjoint extension, is due to Gaffney \cite{gaffney51}.
\end{rem}

\begin{proof}
We show that the quantum time evolution corresponding to $H_0$ maps the set of quantisations of the generators $e_k \otimes h_{U,\xi,g}$ of $C_{\mathcal{R}}(T^\ast \IT^n)$ into itself; since the time evolution consists of a family of automorphisms of C$^\ast$-algebras, the lemma will follow from this.

Let $e_k \otimes h_{U,\xi,g}$ be such a generator.
Note that for each $a \in \IZ^n$, we have
\begin{equation}\label{eq:exp of H0 on eigenvector}
e^{-\frac{itH_0}{\hbar}} \psi_a
= e^{-2\pi^2 i t \hbar \|a\|^2}\psi_a\,.
\end{equation}
Using part (3) of Proposition \ref{prop:Weyl_quantisation_is_well-defined}, we obtain
\begin{align*}
&e^{\frac{itH_0}{\hbar}}\mathcal{Q}^W_\hbar\!\left(e_k\otimes h_{U,\xi,g}\right) e^{-\frac{itH_0}{\hbar}} \psi_a \\
&\quad = e^{2\pi^2 i t \hbar (\|a + k\|^2 - \|a\|^2)} e^{\pi \hbar i(k + 2a) \cdot \xi} g \circ P_U(\pi \hbar(k + 2a)) \psi_{k + a} \\
&\quad = e^{\pi i\hbar (k + 2a) \cdot (\xi + 2 \pi t k)} g \circ P_U(\pi \hbar(k + 2a))\psi_{k + a} \\
&\quad = \mathcal{Q}^W_\hbar \!\left(e_k \otimes h_{U,\tilde{\xi},\tilde{g}}\right) \psi_a\,,
\end{align*}
for each $a \in \IZ^n$, where
\begin{equation*}
\tilde{\xi} := \xi + 2 \pi t P_{U^\perp}(k) \in U^\perp\,,
\end{equation*}
and
\begin{equation*}
\tilde{g} \colon U \rightarrow \IC\,, \quad 
p \mapsto e^{2 \pi i t P_U(k) \cdot p} g(p)\,,
\end{equation*}
is again a Schwartz function on $U$, so $e_k \otimes h_{U,\tilde{\xi},\tilde{g}}$ is a generator of $C_{\mathcal{R}}(T^\ast \IT^n)$.
It follows that the set of generators of $A_\hbar$ is indeed invariant under the free quantum time evolution.
\end{proof}

\begin{rem}
Comparing the proof of Lemma \ref{lem:free quantum time evolution} with the proof of the analogous Lemma \ref{lem:free time evolution}, we see that (for $t=1$) $\tilde{\xi}$ and $\tilde{g}$ are both the same. Indeed, one can easily obtain $$\QW\circ (\Phi^t_0)^*=\tau_t^0\circ\QW\,,$$ which is analogous to a known result for Weyl quantisation on $\R^{2n}$ (proved in higher generality in \cite[Theorem II.2.5.1]{landsman98}). There is generally no such result for non-free time evolution.
\end{rem}
\noindent
In order to deal with the general quantum time evolution, we recall some basic theory about lattices that we need due to the appearance of the lattice $\Z^n$ in $\T^n=\R^n/\Z^n$.
A set of linearly independent vectors $v_1,\ldots,v_l$ in a lattice $\Lambda$ is called \textit{primitive in $\Lambda$} if $\spn_\Z(v_1,\ldots,v_l)=\spn_\R(v_1,\ldots,v_l)\cap\Lambda$.
For instance, every $\Z$-basis of a lattice $\Lambda$ is primitive in $\Lambda$. Furthermore, we have the following result:

\begin{lem}\label{lem: extend primitive set}
	Let $\Lambda \subset \IR^m$ be a lattice.
Every primitive set $v_1,\ldots,v_l$ in $\Lambda$ can be extended to a $\Z$-basis $v_1,\ldots,v_l,v_{l + 1},\ldots,v_m$ of $\Lambda$.
\end{lem}

\begin{proof}
	This is exactly \cite[\textsection 1.3, Theorem 5]{Lekkerkerker69}.
\end{proof}

\noindent 
This will help us prove the main theorem of this section:

\begin{thrm}
Let $V \in C(\IT^n)$.
Then the operator $H = -\frac{\hbar^2}{2} \sum\frac{d^2}{dx_j^2} + M(V)$ with domain $\dom H_0$ (see Lemma \ref{lem:free quantum time evolution}) is self-adjoint.
Let $\big(e^{\frac{-itH}{\hbar}}\big)_{t \in \IR}$ be the corresponding one-parameter group implementing the quantum mechanical time evolution on $L^2(\IT^n)$, and let $(\tau_t)_{t \in \IR}$ be the associated one-parameter group of automorphisms on $B(L^2(\IT^n))$.
Then $(\tau_t)_{t \in \IR}$ preserves $A_\hbar$.
\end{thrm}

\begin{proof}
Self-adjointness of $H$ is a consequence of the Kato--Rellich theorem.
We claim that for each $t \in \IR$, we have
\begin{equation*}
e^{\frac{itH_0}{\hbar}} e^{\frac{-itH}{\hbar}} \in A_\hbar \,.
\end{equation*}
Suppose for the moment that this claim holds true.
Then for each $a \in A_\hbar$ and each $t \in \IR$, we have
\begin{equation*}
\tau_t(a)
= e^{\frac{itH}{\hbar}} a e^{\frac{-itH}{\hbar}}
= \left( e^{\frac{itH_0}{\hbar}} e^{\frac{-itH}{\hbar}} \right)^\ast \tau^0_t(a) \left( e^{\frac{itH_0}{\hbar}} e^{\frac{-itH}{\hbar}} \right).
\end{equation*}
By assumption, the first and the third factors within parentheses are elements of $A_\hbar$, and the second factor is an element of $A_\hbar$ by Lemma \ref{lem:free quantum time evolution}.
It then follows that $\tau_t(a) \in A_\hbar$.

Thus it remains to prove the claim.
As in the proof of \cite[Proposition 6.1]{buchholz08}, we use the fact that the product of two of the elements of the different one parameter groups can be written as a norm-convergent Dyson series, i.e.,
\begin{equation}
e^{\frac{itH_0}{\hbar}} e^{\frac{-itH}{\hbar}}
= \sum_{m = 0}^\infty (i\hbar)^{-m} \int_0^t \int_0^{t_1} \dots \int_0^{t_{m - 1}} \tau^0_{t_1} (M(V)) \cdots \tau^0_{t_m}( M(V) ) \: dt_m \cdots dt_2 \: dt_1.
\label{eq:Dyson_series}
\end{equation}
The integrals in the above expression can be defined in the following way.
First, observe that the function
\begin{equation*}
\IR \rightarrow B(L^2(\T^n)) \,, \quad 
t \mapsto \tau^0_t (M(V)) \,,
\end{equation*}
is bounded and strongly continuous.
It follows that the function
\begin{equation*}
\IR^m \rightarrow B(L^2(\IT^n)) \,, \quad 
(t_1, \ldots, t_m) \mapsto \tau^0_{t_1} (M(V)) \cdots \tau^0_{t_m} (M(V)) \,,
\end{equation*}
is bounded and strongly continuous.
For each $\psi \in L^2(\IT^n)$, one can therefore define the integral
\begin{equation}\label{Dyson term}
\int_0^t \int_0^{t_1} \dots \int_0^{t_{m - 1}} \tau^0_{t_1}( M(V)) \cdots \tau^0_{t_m}(M(V))\psi \: dt_m \cdots dt_2 \: dt_1 \,,
\end{equation}
using Bochner integration, and it is easy to check that the norm of the corresponding operator is less than or equal to $(m!)^{-1} |t|^m \|V\|_\infty^m$, so that the Dyson series is indeed norm-convergent.
As in \cite{buchholz08}, because \eqref{Dyson term} is continuous in $V$ it suffices to prove the claim for potentials $V$ that lie in a dense subset of $C(\T^n)$. If we assume that $V$ is in the span of $\set{e_k}{ k\in\Z^n}$, we can write \eqref{Dyson term} as a sum of relatively explicit expressions.
Thus, we are left to show that for each $t \in \IR$ and each $k_1,\ldots,k_m \in \IZ^n$, the operator
\begin{equation*}
a:=\int_0^t \int_0^{t_1} \dots \int_0^{t_{m - 1}} \tau^0_{t_1} (M(e_{k_1})) \cdots \tau^0_{t_m}( M(e_{k_m}) ) \: dt_m \cdots dt_1 \,,
\end{equation*}
lies in $A_\hbar$. 
A quick computation using \eqref{eq:exp of H0 on eigenvector} gives us
\begin{align*}
	\tau^0_t(M(e_k))\psi_a
&= M(e_k)e^{2\pi^2 i\hbar (\norm{a+k}^2-\norm{a}^2)}\psi_a\\
	&= M(e_k) e^{2\pi^2it\hbar \norm{k}^2}e^{4\pi^2it\hbar k\cdot a}\psi_a\,,
\end{align*}
which shows that, for any $\psi\in L^2(\T^n)$ and $[x]\in\T^n$, we have
\begin{align*}
	(\tau^0_t(M(e_k))\psi)[x] = e^{2\pi ix\cdot k} e^{2\pi^2i\hbar t\norm{k}^2}\psi \left[ x + 2\pi\hbar tk \right] \,.
\end{align*}
Applying this formula many times, we find a function $f_0\in C_b(\R^m)$ that takes values on the unit circle such that
\begin{align*}
	\tau^0_{t_1} (M(e_{k_1})) \cdots \tau^0_{t_m}( M(e_{k_m}) )\psi[x] = e^{2\pi i x\cdot\sum k_i} f_0(t_1,\ldots,t_m)\psi\left[x+2\pi\hbar\sum t_ik_i\right]\,.
\end{align*}
The operator $a$ looks like an integral operator, in the sense that we perform an integral over the variables $t_i$ that appear as $\sum t_i k_i$ in the argument of $\psi$. However, the $k_i$'s may both fail to constitute a linearly independent and a complete set of vectors in $\R^n$. Still, we can relate $a$ to an integral operator, which will be the subject of the rest of the proof. 

We use a special case of Lemma \ref{lem: extend primitive set} (extending an empty primitive set) to find a $\Z$-basis $v_1,\ldots,v_l$ of $\spn_\R(k_1,\ldots,k_m)\cap\Z^n$. Because the $k_i$'s are integral, this is also an $\R$-basis of $\spn_\R(k_1,\ldots,k_m)$. Expressing the $k_i$'s in terms of $v_j$'s as
	$$ k_i = \sum_{j=1}^l c_{ij}v_j\,,$$
we obtain
\begin{align*}
	\psi\bigg[x+2\pi\hbar\sum_{i=1}^m t_i k_i\bigg]&=\psi\bigg[x+2\pi\hbar\sum_{j=1}^l\sum_{i=1}^m t_i c_{ij}v_j\bigg]\\
	&= \psi\bigg[x+2\pi\hbar\sum_{j=1}^l T_0(t_1,\ldots, t_m)_jv_j\bigg]\,,
\end{align*}
for a unique surjective linear map $T_0 \colon \mathbb{R}^m \rightarrow \mathbb{R}^l$. By surjectivity, the map $T_0$ admits a lift to an invertible linear map $T \colon \mathbb{R}^m \rightarrow \mathbb{R}^m$ with respect to the projection $\mathbb{R}^m \rightarrow \mathbb{R}^l$ onto the first $l$ coordinates. Fix such a $T$, and perform a change of variables, replacing $(t_1,\ldots, t_m)$ with $T^{-1}(s)$. We get
\begin{align*}
	a\psi[x] = e^{2\pi i x\cdot\sum k_i}\left|\det T\right|^{-1}\int_K f_0\big(T^{-1}s\big)\psi\bigg[x+2\pi\hbar\sum_{j=1}^l s_j v_j\bigg] \:ds\,,
\end{align*}
for some compact subset $K\subseteq \R^m$. Let $K'$ be the image of $K$ under the projection $\R^k\rightarrow\R^l$ onto the first $l$ coordinates, and define the function $f_1\colon\R^l\rightarrow \C$ by
\begin{align*}
	f_1\colon s_{(1)}\mapsto \left|\det T\right|^{-1}\int_{\R^{m-l}}1_{K}(s_{(1)}\oplus s_{(2)})f_0\big(T^{-1}(s_{(1)}\oplus s_{(2)})\big)\:ds_{(2)}\,.
\end{align*}
One easily finds that $f_1\in L^\infty(\R^l)$. We are now left with the integral
\begin{align*}
	a\psi[x]=e^{2\pi ix\cdot\sum k_i}\int_{K'} f_1(s)\psi\bigg[x+2\pi\hbar\sum_{j=1}^l s_j v_j\bigg]\:ds\,.
\end{align*}
We want to relate the above integral to an integral over the first $l$ components in $\T^n$.
For this purpose, we apply Lemma \ref{lem: extend primitive set} once more to extend $v_1,\ldots,v_l$ to a $\Z$-basis $v_1,\ldots,v_n$ of $\Z^n$, and let $S$ be the matrix whose columns are the vectors $v_1,\ldots,v_n$.
Since $S$ and its inverse are matrices in $GL_n(\Z)$, we find that $\det S=\pm1$.
Moreover, $S$ induces the group automorphism $[x] \mapsto [Sx]$ of $\T^n$, which we can pull back to the unitary map
	\[U\colon L^2(\T^n)\rightarrow L^2(\T^n)\,,\quad U\psi[x]:=\psi[Sx]\,,\]
for which it is straightforward to check (on generators of $A_\hbar$) that $U^{-1} A_\hbar 
U\subseteq A_\hbar$. For $\varphi=\varphi_1\otimes\varphi_2\in L^2(\T^l)\otimes L^2(\T^{n-l})$ we have, denoting $k:=\sum_i k_i$,
\begin{align*}
	U M(e_{-k})aU^{-1}\varphi[x]
	&= \int_{K'} f_1(s)U^{-1}\varphi\bigg[S(x)+2\pi\hbar\sum_{j=1}^l s_j S(e_j)\bigg]\:ds\\
	&=\int_{K'} f_1(s)\varphi\left[x+2\pi\hbar(s\oplus0)\right]\:ds\\
	&= \int_{K'} f_1(s)\varphi_1\left(x_{(1)}+2\pi\hbar s+\Z^l\right)\varphi_2\left(x_{(2)}+\Z^{n-l}\right)\:ds\\
	&=\int_{\T^l} f_2\left(x_{(1)}+\Z^l,s\right)\varphi_1(s)\:ds\:\varphi_2\left(x_{(2)}+\Z^{n-l}\right)\,,
\end{align*}
where $x=x_{(1)}\oplus x_{(2)}$ and $f_2\in L^\infty(\T^l\times\T^l) \subseteq L^2(\T^l\times\T^l)$ denotes the function
\begin{equation*}
f_2(r,s)
:= \sum_{M\in\Z^l} f_1\left(\frac{\iota(s-r)+M}{2\pi\hbar}\right)\,,
\end{equation*}
where $\iota$ denotes the canonical map $\T^l\rightarrow[0,1)^l$.
Note that the above sum has only finitely many nonzero terms since $f_1$ is compactly supported.

In conclusion, we have proved that
\begin{align*}
	a = M(e_{k})U^{-1}(F\otimes\mathbf{1})U\,,
\end{align*}
for an integral operator $F\in L^2(L^2(\T^l))$. By part (3) of Proposition \ref{prop:quantisation_notable_properties}, any compact operator, like $F$, is inside the quantum resolvent algebra on $\T^l\times\R^l$. By part (4) of Proposition \ref{prop:quantisation_notable_properties}, this implies that $F\otimes\mathbf{1}\in A_\hbar$, and hence $U^{-1}(F\otimes\mathbf{1})U\in A_\hbar$. As $M(e_{k})$ is the quantisation of $e_{k}\otimes \unit_{\R^n}$, we find $a\in A_\hbar$. As we have seen, linearity and continuity of the Dyson series imply that $e^{\frac{itH_0}{\hbar}}e^{\frac{-itH}{\hbar}}\in A_\hbar$, and this implies the theorem itself.
\end{proof}

\ifx\mycmd\undefined
	\phantomsection
	\addcontentsline{toc}{section}{References}
	\bibliographystyle{abbrv}
	\bibliography{./../Miscellaneous/References}

	\end{document}
\fi

\phantomsection
\addcontentsline{toc}{section}{References}
\bibliographystyle{abbrv}
\bibliography{./Miscellaneous/References}

% \phantomsection is necessary to rename the bibliography to `References', and to have it appear as such in the table of contents.

\end{document}